\documentclass[a4paper,11pt]{article}
\pdfoutput=1
\usepackage[utf8]{inputenc}
\usepackage{amssymb}
\usepackage{amsmath}
\usepackage{amsthm}
\usepackage{amsfonts}
\usepackage[pdftex]{graphicx}
\usepackage{geometry}
\usepackage{braket}
\usepackage{enumerate}
\usepackage{comment}

\usepackage{subcaption}

\allowdisplaybreaks

\usepackage[table,usenames,dvipsnames]{xcolor}

\usepackage{setspace}
\usepackage[in]{fullpage}
\usepackage{hyperref}
\usepackage{array}
\usepackage{cancel}
\usepackage{wrapfig}
\usepackage[font=small,labelfont=bf]{caption}
\usepackage{pifont}

\usepackage[dvipsnames]{xcolor}
\usepackage{mathtools}

\usepackage{color}

\usepackage{tikz}
\usetikzlibrary{arrows,shapes}
\usetikzlibrary{trees}
\usetikzlibrary{matrix,arrows} 
\usetikzlibrary{positioning}
\usetikzlibrary{calc,through}
\usetikzlibrary{decorations.pathreplacing}
\usepackage{pgffor}	

\numberwithin{equation}{section}

\usepackage[english]{babel}

\usepackage{caption}

\usepackage[nottoc,notlot,notlof]{tocbibind}
\usepackage[nosort]{cite}   
\usepackage{color}

\usepackage{parskip}
\usepackage[T1]{fontenc}

\usepackage{lmodern}
\usepackage{yfonts}

\usepackage{physics}
\usepackage{slashed}
\usepackage{makeidx}
\usepackage[vcentermath]{youngtab}
\usepackage{amsthm}
\usepackage[scr=boondoxo]{mathalfa}

\hypersetup{
	colorlinks=true,
	linktoc=page,
	citecolor=blue,
	linkcolor=blue,
	urlcolor=blue} 

\usepackage{tikz}
\usetikzlibrary{arrows,shapes}
\usetikzlibrary{trees}
\usetikzlibrary{matrix,arrows} 				
\usetikzlibrary{positioning}				
\usetikzlibrary{calc,through}				
\usetikzlibrary{decorations.pathreplacing}  
\usepackage{pgffor}							

\usetikzlibrary{decorations.pathmorphing}	
\usetikzlibrary{decorations.markings}
\tikzset{
   vector/.style={decorate, decoration={snake, amplitude=1pt, segment length=6pt}, draw,double},
   vector2/.style={decorate, decoration={snake, amplitude=1pt, segment length=6pt}, draw},
	provector/.style={decorate, decoration={snake,amplitude=2.5pt}, draw},
	antivector/.style={decorate, decoration={snake,amplitude=-2.5pt}, draw},
    fermion/.style={draw=black, postaction={decorate},
        decoration={markings,mark=at position .55 with {\arrow[draw=black]{>}}}},
    fermionbar/.style={draw=black, postaction={decorate},
        decoration={markings,mark=at position .55 with {\arrow[draw=black]{<}}}},
    fermionnoarrow/.style={draw=black},
    gluon/.style={decorate, draw=black,
        decoration={coil,amplitude=4pt, segment length=5pt}},
    scalar/.style={dashed,draw=black, postaction={decorate},
        decoration={markings,mark=at position .55 with {\arrow[draw=black]{>}}}},
    scalarbar/.style={dashed,draw=black, postaction={decorate},
        decoration={markings,mark=at position .55 with {\arrow[draw=black]{<}}}},
    scalarnoarrow/.style={dashed,draw=black},
    electron/.style={draw=black, postaction={decorate},
        decoration={markings,mark=at position .55 with {\arrow[draw=black]{>}}}},
	bigvector/.style={decorate, decoration={snake,amplitude=4pt}, draw},
}
\tikzset{cross/.style={cross out, draw, 
         minimum size=2*(#1-\pgflinewidth), 
         inner sep=0pt, outer sep=0pt}}

\tikzstyle{block} = [draw, rectangle, 
    minimum height=3em, minimum width=6em]

\newcommand{\sqr}[2]{\lbrack #1 #2 \rbrack}

\newcommand{\cA}{\mathcal{A}}

\newcommand{\cL}{\mathcal{L}}
\newcommand{\cO}{\mathcal{O}}

\usepackage{float}
\usepackage{marvosym}
\usepackage{empheq}
\usepackage{bbold}

\usepackage{tikz}
\usetikzlibrary{shapes}
\usetikzlibrary{arrows}
\usetikzlibrary{positioning}
\usetikzlibrary{decorations.pathmorphing}
\usetikzlibrary{decorations.pathreplacing}
\usetikzlibrary{decorations.markings}

 \def\cA{\mathcal{A}}
 \def\uno{\mbox{1 \kern-.59em {\rm l}}}

\begin{document}


\begin{flushright}
	QMUL-PH-20-19\\
	SAGEX-20-19-E\\
\end{flushright}

\vspace{20pt} 

\begin{center}

	{\Large \bf  From amplitudes to gravitational radiation   }  \\
	\vspace{0.3 cm} {\Large \bf  with cubic interactions  
	and tidal effects}

	\vspace{25pt}

	{\mbox {\sf  \!\!\!\!Manuel~Accettulli~Huber, Andreas~Brandhuber, Stefano~De~Angelis and 				Gabriele~Travaglini{\includegraphics[scale=0.05]{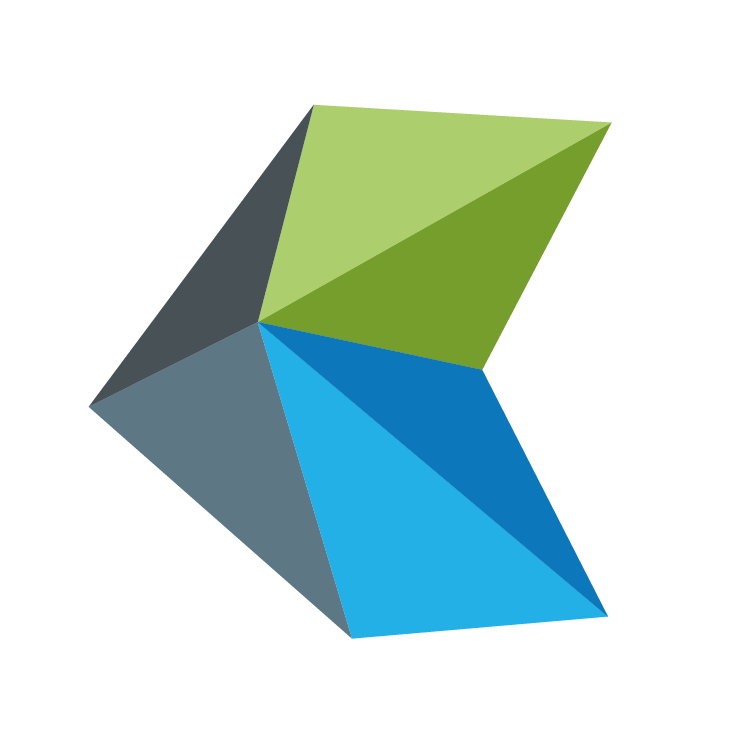}}
	}}
	\vspace{0.5cm}

	\begin{center}
		{\small \em
			Centre for Research in String Theory\\
			School of Physics and Astronomy\\
			Queen Mary University of London\\
			Mile End Road, London E1 4NS, United Kingdom
		}
	\end{center}


	\vspace{40pt}  

	{\bf Abstract}
\end{center}

\vspace{0.3cm}

\noindent
We study the effect of  cubic and tidal interactions on the spectrum of gravitational waves emitted in the inspiral phase of the merger of two non-spinning objects. There are  two independent parity-even cubic interaction terms,  which we take to be $I_1  = {R^{\alpha \beta}}_{\mu \nu} {R^{\mu \nu}}_{\rho \sigma} {R^{\rho \sigma}}_{\alpha \beta}$ and 
$G_3 = I_1-2  R^{\alpha}\,_{\mu}\,^{\beta}\,_{\nu} R^{\mu}\,_{\rho}\,^{\nu}\,_{\sigma} R^{\rho}\,_{\alpha}\,^{\sigma}\,_{\beta}$.
The latter   has vanishing pure graviton amplitudes but modifies mixed scalar/graviton amplitudes which are crucial for our study. 
Working in an effective field theory set-up, we compute the modifications to the quadrupole moment due to  $I_1$, $G_3$  and tidal interactions, from which we obtain the power of gravitational waves radiated in the process to first order in the  perturbations and leading order in the post-Minkowskian expansion. The $I_1$ predictions are novel, 
 and  we  find that our results for $G_3$ are related to the known quadrupole corrections arising from tidal perturbations, although the physical origin of the $G_3$ coupling is unrelated to the finite-size effects underlying tidal interactions. We show this by recomputing such tidal corrections  and by presenting an explicit field redefinition. In the post-Newtonian expansion our results are complete at  leading order, which for the gravitational-wave flux is 5PN for $G_3$ and tidal interactions, and 6PN for $I_1$. Finally, we  compute the corresponding   modifications to the waveforms.

\vfill
\hrulefill
\newline
\vspace{-1cm}
${\includegraphics[scale=0.05]{Sagex.jpeg}}$~\!\!{\tt\footnotesize\{m.accettullihuber, a.brandhuber, s.deangelis, g.travaglini\}@qmul.ac.uk}

\setcounter{page}{0}
\thispagestyle{empty}
\newpage


\setcounter{tocdepth}{4}
\hrule height 0.75pt
\tableofcontents
\vspace{0.8cm}
\hrule height 0.75pt
\vspace{1cm}
\setcounter{tocdepth}{2}

\section{Introduction} 

The first direct detection of gravitational waves and the first observation of a binary black hole merger by the LIGO/Virgo collaboration
\cite{Abbott:2016blz} has opened a new observational window potentially challenging our understanding of gravity.
Anticipating improved experimental sensitivity in the future, high-precision theoretical predictions from general relativity will be required and in the recent few years much  effort went into developing new theoretical tools using traditional and novel approaches. This includes  important   calculations  of the effective gravitational   potential  at  second 
\cite{Damour:1985mt,Gilmore:2008gq}, third \cite{Damour:2001bu, Blanchet:2003gy, Itoh:2003fy,Foffa:2011ub}, fourth \cite{Jaranowski:2012eb, Damour:2014jta,Galley:2015kus,Damour:2015isa, Damour:2016abl, Bernard:2015njp, Bernard:2016wrg, Foffa:2012rn, Foffa:2016rgu,Porto:2017dgs,Porto:2017shd,Foffa:2019yfl,Blumlein:2020pog}, fifth \cite{Foffa:2019hrb,Blumlein:2019zku,Blumlein:2020pyo}  and sixth \cite{Blumlein:2020znm,Bini:2020uiq} post-Newtonian order%
\footnote{For a recent review of the EFT approach to the binary problem \cite{Goldberger:2004jt} in the PN expansion see \cite{Porto:2016pyg}.}, 
as well as in  the post-Minkowskian expansion 
\cite{Bern:2019nnu,Bern:2019crd,Cheung:2020gyp} and  formal developments in computing classical observables from scattering amplitudes 
\cite{Bjerrum-Bohr:2013bxa,Bjerrum-Bohr:2017dxw,Luna:2017dtq,Kosower:2018adc,Guevara:2018wpp,Chung:2018kqs,KoemansCollado:2019ggb,Maybee:2019jus,Guevara:2019fsj,Chung:2019duq,Damgaard:2019lfh,Kalin:2019rwq,Kalin:2019inp,Chung:2020rrz,Cristofoli:2020uzm,Bern:2020gjj,Bern:2020buy,Parra-Martinez:2020dzs,delaCruz:2020bbn,Emond:2020lwi}. 
A related, ambitious question is whether gravitational waves can, now or in the near future, provide feasible tests of modifications of general relativity as implied by string theory or other extensions of  Einstein-Hilbert (EH) gravity. Even if the experimental precision has not been reached today, one can  entertain this tantalising possibility.

An effective field theory (EFT) framework for gravity was advocated in 
 \cite{Donoghue:1994dn}, and  is ideally suited to study systematically higher-derivative corrections to the EH theory. In \cite{Endlich:2017tqa}, this approach was followed  to compute the corrections to the gravitational potential between  compact objects and their effective mass and current quadrupoles due to perturbations quartic in the Riemann tensor, and the corresponding modifications to the waveforms were  then analysed in 
 \cite{Sennett:2019bpc}.  Modifications to the gravitational potential due to cubic interactions in the Riemann tensor were computed in  \cite{Brandhuber:2019qpg, Emond:2019crr} using amplitude techniques, and  the 
 deflection angle and time delay/advance of massless particles of spin $0, 1$ and $2$ were derived in 
\cite{AccettulliHuber:2020oou} for cubic and quartic perturbations in the Riemann tensor as well as for interactions of the type $FFR$\cite{AccettulliHuber:2020oou}.  Terms that are quadratic in the Riemann tensor do not contribute to the classical scattering of particles in four dimensions \cite{Huber:2019ugz}. 
In this paper we wish to  describe dissipative effects in the dynamics of binaries, that  is gravitational-wave radiation,    
from appropriate five-point amplitudes with four massive scalars and one radiation graviton. We perform this study in the presence of cubic  modifications to the EH action and tidal effects. Interestingly, we will see that there is an overlap between these two types of corrections, which are linked by appropriate field redefinitions \cite{deRham:2019ctd,Bern:2020uwk} which we construct explicitly. We note however that the physical origin of these interactions is very different -- for instance, $I_1$  and $G_3$ appear in the low-effective action of bosonic strings, 
or can be induced by integrating out massive matter \cite{Avramidi:1986mj,Avramidi:1990je}.

In the presence of scalars and restricting our focus to parity-even interactions, there are two independent cubic terms: $I_1  := {R^{\alpha \beta}}_{\mu \nu} {R^{\mu \nu}}_{\rho \sigma} {R^{\rho \sigma}}_{\alpha \beta}$ and $I_2 :=  R^{\alpha}\,_{\mu}\,^{\beta}\,_{\nu} R^{\mu}\,_{\rho}\,^{\nu}\,_{\sigma} R^{\rho}\,_{\alpha}\,^{\sigma}\,_{\beta}$. A more natural combination  is in fact $G_3 := \ I_1 \, - \, 2 \, I_2$, which, as is well known, is topological in six dimensions \cite{vanNieuwenhuizen:1976vb} and has vanishing  graviton amplitudes.
In \cite{Camanho:2014apa},  it was argued from studying the  scattering of polarised gravitons
that $I_1$ potentially leads to superluminal effects/causality violation in the propagation of gravitons for impact parameter $b\lesssim \alpha^{\frac{1}{4}}$. Here $\alpha \sim \Lambda^{-4}$ is the coupling constant of the $I_1$ interaction, and  $\Lambda$ is the cutoff of the theory. In that paper,   $\alpha$ was chosen to be  much larger than $G^2\sim M_\text{Planck}^{-4}$. This allows to treat the gravitational scattering in a semiclassical set-up, where predictions can be trusted up to $M_\text{Planck} (> \Lambda)$. 
This question was reinvestigated in an EFT framework in \cite{AccettulliHuber:2020oou}, where it was found that the  $I_1$ interaction leads to a time advance in the propagation of gravitons (but not photons and scalars) when $b\lesssim \alpha^{\frac{1}{4}}$. 
Finally, 
 $G_3$  does not lead to any time advance/delay for massless particles \cite{AccettulliHuber:2020oou}, while still correcting the gravitational potential \cite{Brandhuber:2019qpg, Emond:2019crr}. An identical conclusion for the propagation of massless particles  in the background of a black hole was reached in \cite{deRham:2020ejn}, both for the  $I_1$ and  $G_3$ interactions%
\footnote{Note that for $G_3$ the coefficient $2d_9 + d_{10}$ in  Eq.~(2.24) of \cite{deRham:2020ejn} vanishes.}.

In this respect, an  important observation was made in \cite{deRham:2020zyh}, namely that such superluminality effects (and those observed earlier on in \cite{Drummond:1979pp,Hollowood:2007ku,Goon:2016une}) are unresolvable within the regime of validity of the EFT, and do not lead to violations of causality. In our set-up such violations would indeed occur at $b \lesssim\Lambda^{-1}$, which is at the boundary of the regime of validity of our EFT, while the processes we are interested in only probe the regime where the EFT is valid. 
Above $\Lambda$,  the only known way to restore causality is to introduce  an infinite tower of massive particles \cite{Camanho:2014apa}.
In conclusion,  these observations do not  rule out cubic interactions for our EFT computation, although  they   may impose constraints on the  cutoff -- it  needs to be such that possible effects due to the massive modes, required to ensure  causality, cannot be resolved with current-day experiments. 
We also note that, assuming that these interactions can contribute to any classical gravitational scattering ($\Lambda < M_\text{Planck}$),  then we have $\alpha > G^2$, independently of precise estimates of the cutoff $\Lambda$.

In the following we  work in an effective theory containing cubic and tidal perturbations, and   compute a five-point amplitude  with four massive scalars (representing the black holes) and  one radiated soft graviton. From this,  one can in principle extract   all radiative multipole moments to this order, but for the sake of   our applications we will only  focus on the quadrupole moment induced by the  cubic and tidal interactions, 
from which we  then derive  the corresponding changes 
to the power radiated by gravitational waves and to the waveforms.
Our results for the quadrupole correction are exact to leading order  in the perturbations and in the post-Minkowskian expansion. 
We also take the post-Newtonian expansion of  our results, which  are complete 
at  5PN order for the $G_3$ and   tidal interaction corrections, and at 6PN order for the $I_1$ corrections. We find that the corrections due to $G_3$ have the same form as those generated by a particular type of tidal interaction (although the corresponding coefficients in the EFT action  are  independent). We also explain this result   by   constructing an explicit field redefinition relating the two couplings. 
For the  PN-expanded result of the tidal corrections to the  mass quadrupole we find agreement with \cite{Henry:2019xhg,Marchand:2020fpt,Henry:2020ski}.
The remaining tasks consist in using the corrected  quadrupole moment to compute the modifications compared to EH gravity to the power emitted by the radiated gravitational waves, and the corresponding corrections to the waveforms in the  Stationary Phase Approximation (SPA)%
\footnote{See \textit{e.g.} 
\cite{Damour:1997ub,Buonanno:2009zt} for details of this approximation.}.
Here we follow closely \cite{Sennett:2019bpc}, and also present a comparison with their result obtained with  perturbations that are quartic in the Riemann tensor.

The rest of the paper is organised as follows. In Section \ref{sec:2} we introduce the EFT we are discussing, reviewing some of the relevant results, including the corrections to the gravitational potential from cubic \cite{Brandhuber:2019qpg, Emond:2019crr} 
and tidal interactions  \cite{Bini:2020flp,Kalin:2020mvi,Cheung:2020sdj}. Furthermore,  we point out the vanishing of all graviton amplitudes in  the pure gravity plus $G_3$  theory, and explicitly construct a field redefinition that maps $G_3$ into a tidal perturbation. 
Section \ref{sec:3} contains the  calculation of the relevant four-scalar, one soft graviton amplitude in our EFT, from which we  extract the perturbations to the quadrupole moment. In Section \ref{sec:4} we compute the power radiated by the gravitational waves, and finally in Section \ref{sec:5} the corrections to the waveforms in the SPA. In an Appendix we present some details on the modifications to the  circular orbits due to  the perturbations.

\section{Description of the theory}
\label{sec:2}

\subsection{The EFT action}

We consider an  EFT describing EH gravity with higher-derivative couplings interacting with  two massive scalars. These  model spinless heavy objects, and we also include the leading  tidal interactions in our description which describe finite size effects of the heavy objects. Specifically, the EFT action we consider is  
\begin{equation}
\label{action}
S \ = \ S_{\rm eff}\, + \, S_{\phi_1\phi_2}\, + \, S_{\rm tidal}\ , 
\, 
\end{equation}
where  
\begin{align}
\label{actioneff}
S_{\rm eff}&= \int\!\dd[4] x\sqrt{-g} \,  \bigg[ -\frac{2}{\kappa^2}R    
-  \frac{2}{\kappa^2} \mathcal{L}_6 
- \cdots \bigg] \ 
\end{align}
is the effective action for gravity, with 
 \begin{equation}
\label{L6}
\cL_6  = \frac{\alpha_1}{ 48} \, I_1 + \frac{\alpha_2}{ 24} \, G_3  
\, . 
\end{equation}
 $I_1$ and $G_3$ are the parity-even cubic couplings defined as 
\begin{equation}
  I_1 \,  := \, {R^{\alpha \beta}}_{\mu \nu} {R^{\mu \nu}}_{\rho \sigma} {R^{\rho \sigma}}_{\alpha \beta}\ , 
\qquad G_3 \, := \, I_1 \, - \, 2 \, I_2\, , 
\end{equation}
with 
\begin{align}
I_2 \, := \, R^{\alpha}\,_{\mu}\,^{\beta}\,_{\nu} R^{\mu}\,_{\rho}\,^{\nu}\,_{\sigma} R^{\rho}\,_{\alpha}\,^{\sigma}\,_{\beta}\   . 
\end{align}
The dots in \eqref{actioneff} stand for higher-derivative interactions that we will not consider here. 
The two scalars, with masses $m_1$ and $m_2$, couple to gravity with an action 
 \begin{align}
S_{\phi_1 \phi_2} \  = \int\!\dd[4] x\sqrt{-g} \ \frac{1}{2} \sum_{i=1,2} \left( \partial_{\mu} \phi_i \partial^{\mu}\phi_i - m_i^2 \phi_i^2 \right) \ , 
\end{align}
and in addition we include  higher-derivative  couplings describing tidal effects of extended heavy objects,
\begin{equation}
\label{tidal}
	S_{\rm tidal}= \int\!\dd[4] x\sqrt{-g} \ \frac{1}{4} R_{\mu \alpha \nu \beta} R^{\rho \alpha \sigma \beta}\sum_{i=1,2}\Big(\lambda_i \, \phi_i^2\delta^\mu_\rho \delta^\nu_\sigma \, + \, \frac{\eta_i}{m_i^4}\, \nabla^\mu \nabla^\nu \phi_i \nabla_\rho \nabla_\sigma \phi_i \Big) \, + \, \cdots
	\ .
\end{equation}
These tidal interactions 
were  recently studied in \cite{Cheung:2020sdj}, and the  dots stand for the (Hilbert) series of higher-dimensional operators classified in \cite{Bern:2020uwk,Haddad:2020que}, which will not play any role in this work.
We now briefly discuss  some properties of the interactions  we consider. 

\subsection{Cubic interactions}
The  $I_1$ and $G_3$  interactions   naturally arise in the  low-energy effective description of bosonic string theory, whose  terms cubic in the curvature  can be obtained by making   the replacement  
\begin{align}
\alpha_1 = \alpha_2 \to  \alpha^{\prime \, 2} e^{-4\Phi}
\ 
\end{align}
 in \eqref{L6}, 
where $\Phi$ is the dilaton. 
These interactions are also produced in the process of integrating out massive matter \cite{Avramidi:1986mj,Avramidi:1990je}.
In pure gravity only one of them is independent in four dimensions \cite{lovelock_1970,Edgar:2001vv}, while in the presence of matter coupled to gravity they become independent. 
For the sake of the computation of the power radiated by the gravitational waves performed in later sections we need the correction induced by the cubic interactions to the gravitational potential. The full 2PM computation of this quantity was performed in  \cite{Brandhuber:2019qpg, Emond:2019crr}, and 
expanding their result   one obtains
\begin{align}
\begin{split}
\label{Vofr}
V(\vec{r}, |\vec{p}\, |) & = \, -\frac{G m_1 m_2}{r} \, + \,  \frac{3}{8} \frac{\alpha_1\,  G^2}{r^6}  \frac{(m_1 + m_2)^3}{m_1 m_2} \vec{p}^{\, 2}\,\\
&  -\,  \frac{3}{4}\frac{\alpha_2 \, G^2}{r^6}  m_1 m_2 (m_1 + m_2) \bigg( 1 \, - \, \frac{m_1^2 + m_2^2}{2\, m_1^2 m_2^2}\, \vec{p}^{\, 2}\bigg) + \, \cdots  \ ,
\end{split}
\end{align}
where  the dots indicate higher PN corrections which we do not consider here. Note that the terms proportional to $\alpha_1$ and $\alpha_2$ are the result of a one-loop computation. In the PN expansion, the term proportional to $\alpha_1$ (from the $I_1$ interaction) is  suppressed by a factor of  $\vec{p}^{\, 2}/m_{1,2}^2$ compared to  the dominant correction proportional to $\alpha_2$ (from $G_3$).

\subsubsection*{Amplitudes from the  $G_3$ interaction}
It is well known that, unlike  $I_1$, the $G_3$ interaction 
 has a vanishing three-graviton amplitude and   does not contribute to graviton scattering up to four particles \cite{vanNieuwenhuizen:1976vb, Broedel:2012rc}  --  and in fact to any number of  gravitons. This  can be understood by the fact that  $G_3$ is topological in six dimensions \cite{vanNieuwenhuizen:1976vb}, and therefore  computing 
 tree-level four-dimensional graviton amplitudes from dimensionally reducing the six-dimensional ones  automatically gives zero. Combining this observation with unitarity techniques leads to
\begin{equation}
	\mathcal{M}_{\text{EH} + G_3} (h_1,\ldots , h_n)\big |_{d < 6}= \mathcal{M}_{\text{EH}} (h_1, \ldots , h_n)\big |_{d < 6} \>,
\end{equation}
for any $n$. Hence the $G_3$ interaction does not affect the perturbative dynamics in theories of pure gravity.
However,  if we consider a theory of gravity with matter, {\it e.g.} massive scalars mimicking black holes or neutron stars, the presence of a $G_3$ coupling alters their dynamics.  In particular the four-point amplitude with two gravitons and two scalars  becomes  \cite{Brandhuber:2019qpg,Emond:2019crr}
\begin{equation}
	\label{eq:fourptG3}
	\mathcal{M}_{\text{EH} + G_3}^{(0)} (\phi_1, \phi_2, h^{++}_3 , h^{++}_4) = \mathcal{M}^{(0)}_{\text{EH}} (\phi_1, \phi_2, h^{++}_3 , h^{++}_4) \, + \, i\, \frac{\alpha_2}{32} \left(\frac{\kappa}{2}\right)^2 \sqr{3}{4}^4\, (2m^2+s) \ . 
\end{equation}
The non-trivial contribution to the scattering amplitude of two massive scalars and two gravitons from the $G_3$ interactions modifies the  classical potential in the two-body system, as shown in \cite{Brandhuber:2019qpg,Emond:2019crr}. As we will show below, both $G_3$ and $I_1$ produce corrections to the quadrupole moment already at tree level. Specifically we find that  the $G_3$ quadrupole correction is dominant in the post-Newtonian (PN) expansion, which parallels the results found for the corresponding corrections to the gravitational potential quoted earlier in \eqref{Vofr}.

\subsubsection*{The \texorpdfstring{$G_3$}{G3}  interaction as a tidal effect}

It is easy to show that the contact term proportional to $\sqr{3}{4}^4\, (2m^2+s)$ in the amplitude \eqref{eq:fourptG3} is (up to a numerical coefficient) the amplitude arising from a particular tidal  interactions of the form $ R^{\mu \nu \rho \sigma} R_{\mu \nu \rho \sigma} m^2 \phi^2-\nabla^{\alpha} R^{\mu \nu \rho \sigma} \nabla_{\alpha} R_{\mu \nu \rho \sigma} \phi^2 $. 
This  suggests  that there should exist a four-dimensional field redefinition  mapping the $G_3$ interaction into a tidal effect, as already noticed in \cite{deRham:2019ctd,Bern:2020uwk}%
\footnote{We also observe  that black holes in four dimensions have non-vanishing Love numbers when higher-derivative interactions are considered  \cite{Cardoso:2018ptl,Cai:2019npx}.}.
In this section we  construct  this field redefinition explicitly.  

We begin by rewriting $G_3$ in a more convenient form, making use of two identities in four dimensions \cite{Xu:1987}:
\begin{equation}
\label{3.1} 
	R^{\alpha \beta}\,_{[ \alpha \beta} R^{\mu \nu}\,_{\mu \nu} R^{\rho}\,_{\rho ]} = 0 \> ,
\end{equation}
which translates into
\begin{equation}
\label{3.2} 
	\begin{split}
		R^{\alpha}\,_{\beta} R^{\beta \rho}\,_{\mu \nu} R^{\mu \nu}\,_{\alpha \rho} = \ &\frac{1}{4}{R}^{3}-2R R^{\alpha}\,_{\beta} R^{\beta}\,_{\alpha}+2R^{\alpha}\,_{\beta} R^{\mu}\,_{\nu} R^{\beta \nu}\,_{\alpha \mu}\\
		&-2R^{\alpha}\,_{\beta} R^{\beta}\,_{\mu} R^{\mu}\,_{\alpha}+\frac{1}{4}R R^{\alpha \beta}\,_{\mu \nu} R^{\mu \nu}\,_{\alpha \beta}\, , 
	\end{split}
\end{equation}
and
\begin{equation}
	R^{\alpha \beta}\,_{[ \alpha \beta} R^{\mu \nu}\,_{\mu \nu} R^{\rho \sigma}\,_{\rho \sigma ]} = 0 \> ,
\end{equation}
which, in combination with \eqref{3.2}, leads to
\begin{equation}\label{eq:I2toI1}
	\begin{split}
		R^{\alpha}\,_{\mu}\,^{\beta}\,_{\nu} R^{\mu}\,_{\rho}\,^{\nu}\,_{\sigma} R^{\rho}\,_{\alpha}\,^{\sigma}\,_{\beta}  = \ &  \frac{1}{2}R^{\alpha \beta}\,_{\mu \nu} R^{\mu \nu}\,_{\rho \sigma} R^{\rho \sigma}\,_{\alpha \beta} - \frac{5}{8}{R}^{3}+\frac{9}{2}R R^{\alpha}\,_{\beta} R^{\beta}\,_{\alpha}\\
		& - \frac{3}{8}R R^{\alpha \beta}\,_{\mu \nu} R^{\mu \nu}\,_{\alpha \beta}-3R^{\alpha}\,_{\beta} R^{\mu}\,_{\nu} R^{\beta \nu}\,_{\alpha \mu}+4R^{\alpha}\,_{\beta} R^{\beta}\,_{\mu} R^{\mu}\,_{\alpha} \> .
	\end{split}
\end{equation}
The latter  identity implies  that, in four dimensions,  $G_3$ can be rewritten as
\begin{align}
	\begin{split}
		\left. G_3 \right| _{d=4}  \ &=   \ \frac{3}{4}R \, R^{\alpha \beta}\,_{\mu \nu} R^{\mu \nu}\,_{\alpha \beta}+\frac{5}{4}{R}^{3}-9 R \, R^{\alpha}\,_{\beta} R^{\beta}\,_{\alpha}-8R^{\alpha}\,_{\beta} R^{\beta}\,_{\mu} R^{\mu}\,_{\alpha}+6R^{\alpha}\,_{\beta} R^{\mu}\,_{\nu} R^{\beta \nu}\,_{\alpha \mu}\\[.2em]
		&\sim  \ \frac{3}{4}R\,  R^{\alpha \beta}\,_{\mu \nu} R^{\mu \nu}\,_{\alpha \beta} \>,
	\end{split}
\end{align}
where in the second line we have dropped  all  terms involving more than one  Ricci scalar/tensor. These terms can be traded, via a further field redefinition, for a contact term of the form $R_{\mu \nu \rho \sigma}  \partial^\mu \phi_1 \partial^\nu \phi_2 \partial^\rho \phi_1 \partial^\sigma \phi_2$, which only contributes to quantum corrections to the quadrupole moment. 
Thus
\begin{equation}
	\label{eq:tidal}
	\begin{split}
		S_{\rm eff}&= \int\!\dd[4] x\sqrt{-g} \,  \bigg[ -\frac{2}{\kappa^2}R -  \frac{\alpha_2}{12\,\kappa^2}\, G_3 \bigg] \, + \, S_{\phi_1,\phi_2}\\
		&= \int\!\dd[4] x\sqrt{-g} \,  \bigg[ -\frac{2}{\kappa^2}R -  \frac{\alpha_2}{16\,\kappa^2}\, R\, (R^{\alpha \beta \mu \nu})^2 + \cdots \bigg] + S_{\phi_1,\phi_2}\\
		&\rightarrow \int\!\dd[4] x\sqrt{-g} \,  \bigg[ -\frac{2}{\kappa^2}R + \frac{\alpha_2}{64}\,(R^{\alpha \beta \mu \nu})^2 \sum_{i=1,2} \left(2\, m_i^2 \phi_i^2 -\, \partial_{\mu} \phi_i \partial^{\mu}\phi_i \right) + \mathcal{O}(\alpha_2^{\, 2}) \bigg] \, + \, S_{\phi_1,\phi_2}\ ,
	\end{split}
\end{equation}
where in the last line we have  used the field redefinition
\begin{equation}
	g_{\alpha \beta} \rightarrow g_{\alpha \beta} - \frac{\alpha_2}{32}\, g_{\alpha \beta} R^{\mu \nu}\,_{\rho \sigma} R^{\rho \sigma}\,_{\mu \nu}\ .
\end{equation}
Finally,  integrating by parts and discarding boundary contributions, we can rewrite   
the   new interaction term in \eqref{eq:tidal} as
  \begin{align}
(R^{\alpha \beta \mu \nu})^2 \left(2\, m^2 \phi^2 -\, \partial_{\mu} \phi \partial^{\mu}\phi \right) =  R^{\mu \nu \rho \sigma} R_{\mu \nu \rho \sigma} m^2 \phi^2-\nabla^{\alpha} R^{\mu \nu \rho \sigma} \nabla_{\alpha} R_{\mu \nu \rho \sigma} \phi^2\ , 
\end{align}
 where the second term does not give any classical contribution to the scattering amplitude. 
 Hence,  for the sake of computing classical contributions to amplitudes, we can replace 
 \begin{equation}
	\label{eq:tidalfinal}
	\begin{split}
		S_{\rm eff}&
		\rightarrow \int\!\dd[4] x\sqrt{-g} \,  \bigg[ -\frac{2}{\kappa^2}R + \frac{\alpha_2}{64}\,(R^{\alpha \beta \mu \nu})^2 \sum_{i=1,2} m_i^2 \phi_i^2  + \mathcal{O}(\alpha_2^{\, 2}) \bigg] \, + \, S_{\phi_1,\phi_2}\ , 
	\end{split}
\end{equation}
thereby   explicitly showing that the $G_3$ interaction can be absorbed into  the first 
of the two tidal interactions in \eqref{tidal}.

\subsection{Tidal effects}

During the inspiral phase of binary systems involving at least one extended heavy object like a neutron star, 
corrections due to the finite size of the object(s) increase as the distance between the objects 
decreases. 
These effects can be included systematically using a tidal expansion, \textit{i.e.}~a multipole expansion dominated by the mass quadrupole moment. 
Finite-size effects are  bound to become of  ever increasing importance in the light of future gravitational-wave  experiments, and will  likely  play a key role in a deeper understanding of the internal structure of compact objects. The computation of tidal effects has been addressed in the past by a wide variety of methods, 
recently including complete PM results \cite{Bini:2020flp,Kalin:2020mvi,Cheung:2020sdj} for the conservative dynamics.

In order to compute the modifications to the waveform coming from the tidal interactions in \eqref{tidal} we need 
to expand  the 2PM potential in the conservative Hamiltonian computed in \cite{Bini:2020flp,Kalin:2020mvi,Cheung:2020sdj}  up to 
 $\cO(\vec{p}^{\, 2})$, with the result  
	\begin{align}\label{eq:vtidal}
		\begin{split}
			V_{\rm tidal}(\vec{r}, \vec{p})&=-\frac{3}{2} \frac{G^2}{r^6}\frac{m_2^2}{m_1}\left[8\left(1-\frac{m_1^2 + m_2^2}{2\, m_1^2 m_2^2}\vec{p}^{\, 2}\right)\lambda_1 + \left(1+\frac{2m_1^2+2m_2^2+5m_1 m_2}{m_1^2 m_2^2}\vec{p}^{\, 2}\right) \eta_1\right] \\
		&+ 1\leftrightarrow 2 
		\, + \, \cdots  \ ,
		\end{split}
		\end{align}
	where  the dots indicate higher PN terms.

\section{Quadrupole moments  in  EFTs of gravity}
\label{sec:3}

In the PN framework, the conservative and dissipative dynamics of two objects of mass $m_1$ and $m_2$, coupled to
the gravity effective action \eqref{actioneff}
is described by   the following point-particle effective action \cite{Goldberger:2004jt,Endlich:2017tqa}:
\begin{eqnarray}
\label{correctedL}
S_{\text{pp}}  & = & \int\!dt \ \left[ \frac{1}{2} \mu \, \dot{\vec{r}}^{\, 2} \, - \, V(\vec{r},\vec{p}) \, +\,  \frac{1}{2} Q^{ij}(\vec{r},\vec{p}) R^{0i0j} + \cdots\right] \ ,
\end{eqnarray}
where
\begin{align}
 \label{redmass}
\mu:= \frac{m_1 m_2}{m_1 + m_2}
\ 
\end{align}
 is  the reduced mass,   and  $\vec{r}(t)$ is  the relative position of the two objects. 
 $V\big(\vec{r}, \vec{p}\big)$  denotes    the potential, 
 whose explicit expression to first order in  
 $\alpha_1$,  $\alpha_2$ \cite{Brandhuber:2019qpg, Emond:2019crr}, and 
 $\lambda_{1,2}$,  $\eta_{1,2}$  \cite{Bini:2020flp,Kalin:2020mvi,Cheung:2020sdj} 
 is obtained by summing \eqref{Vofr} and \eqref{eq:vtidal}, 
 and $Q^{ij}\big(\vec{r}, \vec{p}\big)$ is the quadrupole moment, to be computed below. 
 The dots represent higher-order terms that will be irrelevant in our analysis. 
This action can be trusted in the inspiral phase before the objects reach relativistic velocities.

We now present the computation of the five-point amplitude  $\phi_1 \phi_2 \to \phi_1 \phi_2 + \overline{h}(k)$ with four scalars and one radiated soft graviton $\overline{h} (k)$. Its momentum $k^\mu$ is on shell, while the momentum of the graviton exchanged between the two objects is purely spacelike (corresponding to an instantaneous interaction), and in our set-up is given by $q^\mu= -p_1^\mu-p_2^\mu = (0, \vec{q})$. Furthermore, the energy of the radiated graviton is such that 
$k^0 \ll |\vec{q}\,|$, so that $k^\mu$ can be ignored
for practical purposes, and the radiated graviton enters the amplitude only through its associated Riemann curvature tensor $\overline{R}_{\alpha\beta\mu\nu}$. Finally, because we are only interested in classical contributions ({\it i.e.} $\cO (\hbar^0)$), we keep only the leading terms in $\vec{q}^{\, 2}$.

In the following we first compute fully relativistic scattering amplitudes and then perform the PN expansion to extract the correction to the quadrupole term in the effective action (\ref{correctedL}).
In  the centre-of-mass frame, the momenta of the particles  can be parametrised as 
\begin{align}
\begin{split}
\label{kinematics}
p_1^\mu & =  -\Big(E_1,\,  \vec{p}-\frac{\vec{q}}{2}\Big) \, ,  \qquad 
p_4^\mu \ = - \Big(E_4,\, -\vec{p}+\frac{\vec{q}}{2}\Big) \, , \\
p_2^\mu & =  \Big(E_2,\, \vec{p}+\frac{\vec{q}}{2}\Big) \, ,  \qquad
\ \ \, p_3^\mu \ =  \Big(E_3,\, -\vec{p}-\frac{\vec{q}}{2}\Big)\ ,
\end{split}
\end{align}
with $p_1^2 = p_2^2 = m_1^2$, $p_3^2 = p_4^2 = m_2^2$. Furthermore,  we have 
\begin{align}
\label{ener}
\begin{split}
E_1 \,=\,E_2\,=\,\sqrt{m_1^2 +\vec{p}^{\, \,2}+\vec{q}^{\, \,2}/4}\ , \qquad 
E_3\, =\,E_4\,=\, \sqrt{m_2^2 +\vec{p}^{\, \, 2}+\vec{q}^{\, \,2}/4}\ , 
\end{split}
\end{align}
 where  $\vec{p} \, \cdot \, \vec{q}=0$ because  of momentum conservation. In our all-outgoing convention for the external lines, the four-momenta $p_1$ and $p_4$ correspond to the incoming particles, and hence   their energies are negative.

\begin{figure}
	\centering
	\begin{tikzpicture}[scale=15,baseline={([yshift=-1mm]gr.base)}]
	
		\def\x{0}
				\def\y{0}

				\node at (-5.5pt+\x,-2.5pt+\y) {$\mathcal{A}_{\mathcal{O}}$};
				\node at (-4.2pt+\x,-2.5pt+\y) {$\equiv$};
	
				\node at (-3pt+\x,0+\y) (sc1) {$\phi_1(p_1)$};
				\node at (3pt +\x,0+\y) (sc2) {$\phi_1(p_2)$};
				\node at (0+\x,-2.5pt+\y) (gr) {};
				\node at (3pt+\x,-2.5pt+\y) (gr2) {$\mu \nu \rho \sigma$};
				\node at (-3pt+\x,-5pt+\y) (sc3) {$\phi_2(p_4)$};
				\node at (3pt +\x,-5pt+\y) (sc4) {$\phi_2(p_3)$};
				\node at (-0.7pt+\x,-2.5pt+\y) {$\mathcal{O}$};
				\node at (7pt+\x,-2.35pt+\y) {$\times \overline{R}^{\mu \nu \rho \sigma}(k)\Big|_{k\to 0}$};
				\node at (0.7pt+\x,-1.2pt+\y) {$q$};

				\draw[thick,double] (sc1) -- (sc2);
				\draw[thick,double] (sc3) -- (sc4);
				\draw[vector] (0+\x,0+\y) -- (0+\x,-5pt+\y);
				\draw[vector] (gr.center) -- (gr2);
				\draw [->,>=stealth] (0.3pt+\x,-0.3pt+\y) -- (0.3pt+\x,-2.1pt+\y);
\end{tikzpicture}
\caption{The single diagram contributing to the radiation process with an insertion of the operators $\mathcal{O}=I_1,I_2$. All momenta are treated as outgoing and the radiated graviton is taken to be soft.}
\label{fig:radiation}
\end{figure}

\subsection{The amplitude with cubic interactions}

Our next task is to  compute the  five-point amplitude    $\mathcal{A}_{\mathcal{O}}$ shown in Figure~\ref{fig:radiation},  with $\mathcal{O}=I_1,I_2$ (which we can then combine to obtain $\cA_{G_3}$). 
 We first obtain its relativistic expression, factoring out a single Riemann
tensor associated with the radiated graviton, and  then split the Lorentz indices into time  and 
spatial components and isolate the terms contracted into $\overline{R}_{0i0j}$. Upon Fourier transforming to position space, these components will allow to directly read off $Q_{ij}$ by matching to the Hamiltonian density associated to the point particle effective action \eqref{correctedL}.
The classical  relativistic results are, for $I_1$:
\begin{align}
\begin{split}
\mathcal{A}_{I_1} \, = \, i\,  (\alpha_1 + 2 \alpha_2) 
 \,  \left( \frac{\kappa}{2}\right)^2 \,  \dfrac{q^\mu q^\rho}{q^2} \Big[ m_1^2  \, p_3^\nu   p_3^\sigma + 
m_2^2 \, p_1^\nu   p_1^\sigma - 2 (p_1 \cdot p_3)  p_1^\nu  p_3^\sigma\Big] \overline{R}_{\mu \nu \rho \sigma}
\ , 
\end{split}
\end{align}
while for $I_2$:
\begin{align}
\mathcal{A}_{I_2}\, = \, \frac{i}{2} \alpha_2 \left( \frac{\kappa}{2} \right)^2 \frac{q^\mu q^\rho }{q^2}  
\big( m_1^2  \, p_{3}^{\nu} p_{3}^{\sigma}\, + \,  m_2^2 \,  p_{1}^{\nu} p_{1}^{\sigma}  \big) \overline{R}_{\mu \nu\rho \sigma}
\ . 
\end{align}
Note that the result for the  $G_3$ interaction  introduced in \eqref{L6}  can be obtained as
\begin{align}
 \mathcal{A}_{G_3} \ := \ (\mathcal{A}_{I_1} + \mathcal{A}_{I_2})\big|_{\alpha_1=0} \ .
 \end{align}
The terms in the amplitude contributing to the quadrupole radiation are then
\begin{equation}\label{eq:AI1}
	\mathcal{A}_{I_1}(q)=-i(\alpha_1+2\alpha_2)\left( \frac{\kappa}{2} \right)^2\left(m_1^2 E_4^2+m_2^2E_1^2 -2E_1^2 E_4^2 - 2\vec{p}^{\, 2}E_1 E_4\right)\dfrac{q^i q^j}{\vec{q}^{\, 2}} \, \overline{R}_{0i0j} + \cdots \> ,
\end{equation}
and
\begin{equation}\label{eq:AI2}
	\mathcal{A}_{I_2}(q)=-i\frac{\alpha_2}{2}\left( \frac{\kappa}{2} \right)^2\left(m_1^2 E_4^2+m_2^2 E_1^2\right)\dfrac{q^i q^j}{\vec{q}^{\, 2}} \, \overline{R}_{0i0j} +\cdots\> ,
\end{equation}
where we have used  that $E_3=E_4$ in order to write the result as a function of the energies and momenta of the incoming particles $p_1$ and $p_4$. The dots  stand for  additional terms proportional to $\overline{R}_{0ijk}$ and $\overline{R}_{ijkl}$, which can also be extracted from our result.

\subsection{The amplitude with tidal effects}
A calculation similar to the one outlined in the previous section leads to the fully relativistic result 
\begin{equation}
	\begin{split}
		\mathcal{A}_{\mathrm{tidal}} (q) &= i \left(\frac{\kappa}{2}\right)^2 \frac{q^{\mu} q^{\rho}}{q^2} \,\bigg\{ 8 \lambda_{1}\, p_{4}^{\nu}\, p_{4}^{\sigma} + 8 \lambda_{2}\, p_{1}^{\nu}\, p_{1}^{\sigma}\\ 
		&+\frac{1}{2}\,\left[(m_1^2 + m_2^2 - t)^2 -2 m_1^2\, m_2^2 \right] \left(\frac{\eta_{2}}{m_2^4}\, p_{4}^{\nu}\, p_{4}^{\sigma} + \frac{\eta_{1}}{m_1^4}\, p_{1}^{\nu}\, p_{1}^{\sigma} \right) \bigg\} \overline{R}_{\mu \nu \rho \sigma} \ ,
	\end{split}
\end{equation}
which, upon expanding in the spatial and time components, reads
\begin{equation}
\label{eq:Atidal}
	\begin{split}
		\mathcal{A}_{\mathrm{tidal}} (q) &= - i \left(\frac{\kappa}{2}\right)^2 \bigg\{ 8 \lambda_{1}\, E_4^2 + 8 \lambda_{2}\, E_1^2\\ 
		&+\left[2(E_1 E_4 + \vec{p}^{\,2})^2 - m_1^2\, m_2^2\right] \left(\eta_{2} \frac{E_4^2}{m_2^4} + \eta_{1} \frac{E_1^2}{m_1^4} \right) \bigg\} \frac{q^{i} q^{j}}{\vec{q}^{\,2}} \overline{R}_{0 i 0 j} + \cdots
		\ , 
	\end{split}
\end{equation}
where the ellipses stand once again for terms proportional to $\overline{R}_{0ijk}$ and $\overline{R}_{ijkl}$ which we will not need in the remainder of this paper.

\begin{figure}
	\centering
	\begin{tikzpicture}[scale=15,baseline={([yshift=-1mm]gr.base)}]
	
		\def\x{0}
				\def\y{0}


				\node at (-7.5pt+\x,-2.5pt+\y) {$\mathcal{A}^{\mathcal{O}}_{\mu \nu \rho \sigma}$};
				\node at (-6pt+\x,-2.5pt+\y) {$\equiv$};
	
				\node at (-4pt+\x,0+\y) (sc1) {$\phi_1(p_1)$};
				\node at (4pt +\x,0+\y) (sc2) {$\phi_1(p_2)$};
				\node at (0+\x,-2.5pt+\y) (gr) {};
				\node at (4pt+\x,-2.5pt+\y) (gr2) {$\mu \nu \rho \sigma$};
				\node at (-4pt+\x,-5pt+\y) (sc3) {$\phi_2(p_4)$};
				\node at (4pt +\x,-5pt+\y) (sc4) {$\phi_2(p_3)$};
				\node at (0+\x,-5.8pt+\y) {$\mathcal{O}$};
				\node at (0.7pt+\x,-2.5pt+\y) {$q$};

				\draw[thick,double] (sc1) -- (sc2);
				\draw[thick,double] (sc3) -- (sc4);
				\draw[vector] (0+\x,0+\y) -- (0+\x,-5pt+\y);
				\draw[vector] (0+\x,-5pt+\x) -- (gr2.west);
				\node at (0+\x,-5pt+\y) [circle, fill=black, inner sep=2.5pt]{};
				\draw [->,>=stealth] (0.3pt+\x,-1.5pt+\y) -- (0.3pt+\x,-3.5pt+\y);

				\node at (6.5pt+\x,-2.5pt+\y){ +};

				\def\x{13pt}
				\def\y{0}

	
				\node at (-4pt+\x,0+\y) (sc1) {$\phi_1(p_1)$};
				\node at (4pt +\x,0+\y) (sc2) {$\phi_1(p_2)$};
				\node at (0+\x,-2.5pt+\y) (gr) {};
				\node at (4pt+\x,-2.5pt+\y) (gr2) {$\mu \nu \rho \sigma$};
				\node at (-4pt+\x,-5pt+\y) (sc3) {$\phi_2(p_4)$};
				\node at (4pt +\x,-5pt+\y) (sc4) {$\phi_2(p_3)$};
				\node at (0+\x,0.8pt+\y) {$\mathcal{O}$};
				\node at (0.7pt+\x,-2.5pt+\y) {$q$};

				\draw[thick,double] (sc1) -- (sc2);
				\draw[thick,double] (sc3) -- (sc4);
				\draw[vector] (0+\x,0+\y) -- (0+\x,-5pt+\y);
				\draw[vector] (0+\x,0+\y) -- (gr2.west);
				\node at (0+\x,0+\y) [circle, fill=black, inner sep=2.5pt]{};
				\draw [->,>=stealth] (0.3pt+\x,-1.5pt+\y) -- (0.3pt+\x,-3.5pt+\y);

			
\end{tikzpicture}
\caption{The two diagrams contributing to the gravitational radiation, where $\cO$  denotes any of the two tidal interactions in \eqref{tidal}.   An overall Riemann tensor of the radiated graviton is
 factored out, so that  $\mathcal{A}^{\mathcal{O}}=\mathcal{A}^{\mathcal{O}}_{\mu \nu \rho \sigma}\overline{R}^{\mu \nu \rho \sigma}(k\to 0)$.}
\end{figure}
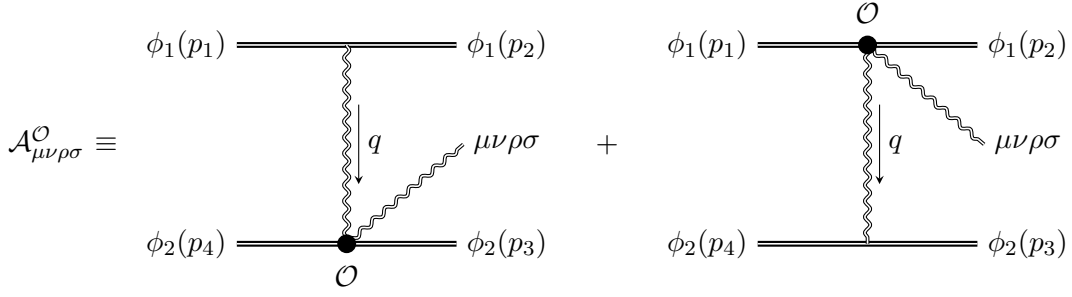

\subsection{The quadrupole corrections}

Next we extract the corrections to the mass quadrupole moment $Q_{ij}$ from \eqref{eq:AI1}, \eqref{eq:AI2} and \eqref{eq:Atidal}. To do so we simply match the appropriately normalised and Fourier-transformed $\mathcal{A}_{\mathcal{O}}$, as defined in \eqref{eq:quadmatch} below, to the quadrupole contribution in \eqref{correctedL}%
\footnote{For further details on the procedure see for example \cite{Goldberger:2004jt,Endlich:2017tqa}.}.
To begin with, we perform the relevant Fourier transforms using
\begin{equation}
	\begin{split}
		\displaystyle\int dt \int \frac{d^3q}{(2\pi)^3} \frac{q_i q_j}{|\vec{q}\,|^{\, 2}} e^{i \vec{q}\cdot\vec{r}} \, \overline{R}^{0i0j} 
	& =-\frac{3}{4\pi}\displaystyle\int dt \frac{1}{r^5}\left(x_i x_j -\frac{1}{3} r^2 \delta_{ij}\right)\overline{R}^{0i0j}\, . 
	\end{split}
\end{equation} 
Taking into account the non-relativistic normalisation factor of $-i/4E_1E_4$, we arrive at the quadrupole-like terms
\begin{equation}\label{eq:quadmatch}
	\begin{split}
		\widetilde{\mathcal{A}}_{\mathcal{O}}^{\rm quad}(r):&=-i \, \frac{\mathcal{A}_{\mathcal{O}}^{\rm quad}(r)}{4E_1E_4} \\
		&=\frac{1}{2} \, C_{\mathcal{O}} (E_i,m_i,\vec{p}^{\, 2})\, \displaystyle\int dt \frac{1}{r^5}\left(x_i x_j -\frac{1}{3} r^2 \delta_{ij}\right)\overline{R}^{0i0j}\, , 
	\end{split}
\end{equation}
where $C_{\mathcal{O}}$ are coefficients depending on the energies and masses as well as $\vec{p}^{\, 2}$ of the heavy particles, with
\begin{equation}\label{eq:coefs}
	\begin{split}
		C_{I_1}(E_i,m_i,\vec{p}^{\, 2})&=\frac{3}{8\pi}\, (\alpha_1+2\alpha_2) \,\left( \frac{\kappa}{2}\right)^2\left(m_1^2 \frac{E_4}{E_1}+m_2^2 \frac{E_1}{E_4}-2E_1E_4-2\vec{p}^{\, 2}\right) \> ,\\
		C_{I_2}(E_i,m_i,\vec{p}^{\, 2})&=\frac{3}{16 \pi}\alpha_2 \, \left( \frac{\kappa}{2}\right)^2 \left(m_1^2\frac{E_4}{E_1}+m_2^2 \frac{E_1}{E_4}\right)
		 \> ,\\
	C_{\rm tidal}(E_i,m_i,\vec{p}^{\, 2})&=	\frac{3}{8\pi} \, \left( \frac{\kappa}{2}\right)^2 \left\{8\lambda_1 \, \frac{E_4}{E_1} +8\lambda_2 \, \frac{E_1}{E_4} \right. +\\
	& \hspace{0.2cm} \left. \left[2\big(E_1E_4+\vec{p}^{\, 2}\big)^2-m_1^2 m_2^2\right]\left(\eta_1 \, \frac{E_1}{E_4 m_1^4}+\eta_2 \, \frac{E_4}{E_1 m_2^4}\right) \right\}
		  \>.
	\end{split}
\end{equation}
Comparing \eqref{eq:quadmatch} with the Hamiltonian density obtained from the action \eqref{correctedL},  
we conclude that the modifications to the quadrupole moment arising from the cubic and tidal  couplings  are given by
\begin{equation}
	Q^{ij}_{\mathcal{O}}\, =\, \frac{C_{\mathcal{O}}}{\mu \, r^5}Q^{ij}_N 
	\ , 
\end{equation}
where we have introduced the leading-order  quadrupole moment in the EH theory for a binary system with masses $m_1$ and $m_2$,\begin{equation}
\label{Newtonqua}
	Q^{ij}_{N}\, =\, \mu\left(x^i x^j -\frac{1}{3} r^2 \delta^{ij}\right) \>, 
\end{equation}
with  $\mu$ being the reduced mass defined in \eqref{redmass}. 
Combining the various correction terms, we arrive at
\begin{equation}
\label{eq:QPN}
	\begin{split}
		Q^{ij}=Q^{ij}_N+Q^{ij}_{I_1}+Q^{ij}_{I_2} +Q^{ij}_{\rm tidal}=\left(1+\frac{C_{I_1}}{\mu \, r^5}+\frac{C_{I_2}}{\mu \, r^5}+ \frac{C_{\rm tidal}}{\mu \, r^5}\right)Q^{ij}_N \> .
	\end{split}
\end{equation} 
It is interesting to write  the three coefficients $C_{I_1}$, $C_{I_2}$ and  $C_{\rm tidal}$  in the PN expansion. Keeping terms up to first order in $\vec{p}^{\, 2}$ one has
\begin{equation}\label{eq:expandedquad}
	\begin{split}
	C_{I_1}^{\rm PN}&=-3 \, G\, \big(\alpha_1+2\alpha_2\big) \, M \frac{\vec{p}^{\, 2}}{\mu}	  \ , \\
	C_{I_2}^{\rm PN}&= \, 3\, G \, \alpha_2 \, m_1 m_2    \ , \\
	C_{\rm tidal}^{\rm PN}&= \, 3 \, G \left[8\lambda_1 +\eta_1+\frac{1}{2M}\Big( 8(m_1-m_2)\lambda_1+(3m_1+5m_2)\eta_1\Big)\frac{\vec{p}^{\, 2}}{\mu^2} \right] \frac{m_2}{m_1} \\
		&+1\leftrightarrow 2 \> ,
\end{split}
\end{equation}
where 
\begin{align}
M:=m_1+m_2 \ , 
\end{align}
and, as usual,   $\kappa^2:=32\pi \, G$. 
For convenience we also quote   the contribution due    to the $G_3$ interaction alone -- this  is given by 
\begin{equation}
	Q^{ij}_{G_3}=\Big( Q^{ij}_{I_1}+Q^{ij}_{I_2}\Big)\bigg|_{\alpha_1=0}\, = \ 3\, G \, \alpha_2 \, \frac{M}{r^5}\left(1-\frac{2\vec{p}^{\, 2}}{\mu^2}\right)Q^{ij}_N \ . 
\end{equation}

\section{Power radiated by the gravitational waves}
\label{sec:4}

\newcommand{\ro}{r_\circ}
\newcommand{\rn}{r_N}
\newcommand{\rd}{\delta r}

We can now compute the power radiated by the gravitational waves in the approximation of circular orbits.
In the EH theory, the radius of the circular orbit is given  by the well-known formula
\begin{align}
\label{erreenne}
\rn =  \left( \frac{G M}{\Omega^2} \right)^{\frac{1}{3}} \ . 
\end{align}
In the presence of the cubic and tidal interactions, this quantity gets  modified as
\begin{equation}\label{errezero}
	\begin{split}
		\ro \, &= \, \rn + \rd \> ,\\[.2em]
		\rd \,& = \, \Omega^3 \left[-\frac{\alpha_1}{2}v + \left(\frac{\alpha_2}{2}+8\lambda_{12}\right)\left(\frac{3}{v}+v(2\nu -1)\right) +\eta_{12}\left(\frac{3}{v}+v(\nu+2)\right)\right]  \, + \, \cO(g_i^2)\ ,
	\end{split}
\end{equation}
where $g_i$ stands for any of the coupling constants of the cubic and tidal perturbations. We also introduced the symmetric mass ratio $\nu$ defined as
\begin{align}
\nu \, := \, \frac{m_1 m_2}{M^2} 
\ ,
\end{align}
and the parameter
\begin{align}
v \, := \rn \Omega = \, (GM \Omega)^\frac{1}{3}\  ,
\end{align}
as well as the following combinations of the couplings
\begin{equation}\label{eq:newcoup}
	\lambda_{12}:=\mu \left(\frac{\lambda_1}{m_1^3}+\frac{\lambda_2}{m_2^3}\right) \> , \hspace{0.3cm}
	\eta_{12}:=\mu \left(\frac{\eta_1}{m_1^3}+\frac{\eta_2}{m_2^3}\right) \> .
\end{equation}
Finally,  $\Omega$ denotes the angular velocity on the circular orbit, and the value $\rd$ has been computed using \eqref{errezero} and \eqref{eq:vprime}, where  the potentials entering \eqref{eq:vprime} are given in \eqref{Vofr} and \eqref{eq:vtidal}.
The total energy per unit mass $M$ of the system, to first order in the couplings, is then given by
\begin{equation}\label{bindingE}
	\begin{split}
			E (v) & = \  -\frac{1}{2}\nu  \, v^2 + \frac{9}{4}\frac{v^{12}}{(GM)^4} \nu \left(\alpha_2 +16 \lambda_{12}+2\eta_{12}\right)
			+ \frac{11}{8}\frac{v^{14}}{(GM)^4}\Big[-\nu \alpha_1
			\\ & 
			+\nu (2\nu -1)\left(\alpha_2+16\lambda_{12}\right)+4\nu(\nu+2)\eta_{12}\Big] 		\ .
	\end{split}
\end{equation}
The above formula is complete at leading order in all of the perturbations (that is $\cO(v^{12}$) and at 
$\cO(v^{14})$ for the $\alpha_1$ correction only. The remaining  $\cO(v^{14})$ terms  have been obtained from a small-velocity expansion of our 2PM result, and in order to get a complete result at that PN order   one would  need to include also  the 3PM corrections to the potential generated by cubic and tidal interactions%
\footnote{Similar considerations  apply to our results 
 for the flux in \eqref{bike}.}. 
 We have also  compared the contribution to the energy from the $\eta_{1,2}$ corrections to \cite{Henry:2020ski}, finding agreement (after mapping their coefficients $\mu_A^{(2)}$ to ours)%
 \footnote{For further details on mapping field-theory  to point-particle  actions  see \textit{e.g.} \cite{Chung:2018kqs,Kim:2020dif}}.
  
Next, we compute  the leading-order gravitational-wave  flux using the  quadrupole formula 
\begin{align}
\label{FNN}
\mathcal{F}(v) \ = \ \frac{G}{5} \langle \dddot{Q}^{ij}\ \dddot{Q}^{ij}\rangle \ , 
\end{align}
using the result of our computation for $Q^{ij}$ in  \eqref{eq:QPN}. To  first order in  the couplings $\alpha_1$ and  $\alpha_2$ the flux becomes 
 \begin{align}
	\begin{split}
	\label{effediv}
   \mathcal{F}(v) 
	&= \frac{G}{5} \langle\dddot{Q}^{ij}_N \  \dddot{Q}_{N}^{ij}\rangle  \left[ 1\, +\, \frac{2}{\mu r^5}\left(C_{I_1}^{\rm PN} + C_{I_2}^{\rm PN}+C_{\rm tidal}^{\rm PN}\right)\right] \, + \, \cO(\alpha_i^2) \, , 
	\end{split}
	\end{align}
where the PN-expanded coefficients $C_{\mathcal{O}}^{\rm {PN}}$ are explicitly given  in \eqref{eq:expandedquad}.

Two comments are in order here. First, we note  that the prefactor $\langle\dddot{Q}^{ij}_N  \dddot{Q}_{N}^{ij}\rangle $ is evaluated on the  radius $\ro$ of the circular orbit in the presence of the cubic and tidal interactions, as given in  \eqref{errezero}. 
Furthermore,  the quantity  $\vec{p}^{\, 2}:=p_r^2+{p_\phi^2}/{r^2}$ can be obtained using the fact that $p_r=0$ on the circular orbit while  $p_\phi:=l$ is a constant, which can be determined  from  Hamilton's equations, with the result
\begin{align}
\label{ellefunctionofomega}
l \ := \  \frac{\mu \ro^2\Omega }{1 + 2 \mu U(\ro)} \ , 
\end{align}
where $r_\circ$ is given in \eqref{errezero} and $U(r)$ is the part of the potential proportional to $\vec{p}^{\, 2}$, following the conventions of Appendix \ref{app:A}. Using these relations, $\vec{p}^{\, 2}$ is re-expressed as a function of $\Omega$, the masses,  and the couplings.

Factoring out the standard  power radiated by the gravitational wave in 
EH,
\begin{align}
\label{FNNbis}
\mathcal{F}_{N}(v) := \frac{G}{5} \left. \langle \dddot{Q}^{ij}_N\ \dddot{Q}_{N}^{ij}\rangle\right|_{r=r_N} \ = \ \frac{32}{5}  G\mu^2  r_N^4 \Omega^6  = \frac{32}{5} \frac{\nu ^2 v^{10}}{G}
 \ ,
\end{align}
we can rewrite the expression for the flux as
\begin{equation}
	\begin{split}
	\label{bike}
		\mathcal{F}(v) &= \frac{32}{5} \frac{\nu ^2 v^{10}}{G} \left[1+\frac{v^{10}}{(GM)^4}\left(12\, \alpha_2+144 \,\lambda_{12}+48 \, \lambda_{12}'+ 18\eta_{12}+6\eta_{12}'\right) \right. \\
		&\left. +\frac{v^{12}}{(GM)^4}\Big[ -8\, \alpha_1+2(2\nu -7)\, \alpha_2+8(8\nu-7)\, \lambda_{12}+24 \, \lambda_{12}'+(8\nu +31)\, \eta_{12}+9\, \eta_{12}'\Big]\right] \> ,	
	\end{split}
\end{equation}
with $\lambda_{12}$ and $\eta_{12}$ defined in \eqref{eq:newcoup} and
\begin{equation}
	\lambda_{12}':=\frac{1}{M} \left(\frac{\lambda_1}{m_1}+\frac{\lambda_2}{m_2}\right) \> , \hspace{0.3cm}
	\eta_{12}':=\frac{1}{M}  \left(\frac{\eta_1}{m_1}+\frac{\eta_2}{m_2}\right) \> .
\end{equation}
Similarly to \eqref{bindingE}, the first line and the $\alpha_1$ term in the second line   of   \eqref{bike} 
 are complete.
 We also note that the $\eta_{1,2}$ part of the tidal flux is  in agreement with  \cite{Henry:2020ski}.

\section{Waveforms in EFT of gravity}
\label{sec:5}

Following \cite{Sennett:2019bpc} we can also compute the correction induced by the cubic and tidal interactions to the gravitational phase in the saddle point approximation. In this approach, the waveform in the frequency domain is written as%
\footnote{See for example Section III F of  \cite{Buonanno:2009zt} for a detailed derivation.}
\begin{align}
\tilde{h}_{\rm SPA} (f) \sim \exp\Big [ i \Big(  \psi_f(t_f) - \frac{\pi}{4}\Big) \Big]
\ , 
\end{align}
where 
\begin{align}
\psi(t) := 2 \pi f t - \phi (t)\ . 
\end{align}
Here $\phi(t)$ is the orbital phase, while $\dot{\phi}(t) = \pi F(t)$ defines the instantaneous frequency $F(t)$ of the gravitational wave. $t_f$ is defined as the time where 
\begin{align}
\left.\dot{\psi}(t) \right|_{t=t_f} \ = \ 0
\ , 
\end{align}
implying that $F(t_f) = 2 f $. 
In the adiabatic approximation, the work of \cite{Buonanno:2009zt,Damour:1997ub} provides explicit formulae for $ \psi_{\rm SPA}(t_f)$ and $t_f$: 
\begin{align}
\label{psiSPA}
 \psi_{\rm SPA} (t_f) & = \ 2 \pi f t_{\rm ref} - 2 \phi_{\rm ref} + \frac{2}{G} \int_{v_f}^{v_{\rm ref}}\!dv \  (v_f^3 - v^3) \frac{E^\prime(v)}{\mathcal{F}(v)}  \ , \\
 \label{tf}
t_f & = \ t_{\rm ref} +M  \int_{v_f}^{v_{\rm ref}}\!dv \  \frac{E^\prime(v)}{\mathcal{F}(v)}  \ , 
 \end{align}
where $v_{\rm ref} = v(t_{\rm ref})$  and $t_{\rm ref}$ are integration constants,   $v_{f } :=  (\pi G M f)^{\frac{1}{3}}$, and 
$E(v)$  and $\mathcal{F}(v)$ were  computed to lowest order in the cubic and tidal  perturbations in \eqref{bindingE} and \eqref{effediv}, respectively. 

We can now compute the correction to  $\psi_{\rm SPA} (t_f) $ due to the presence of the perturbations, expanding the ratio $E'(v)/\mathcal{F}(v)$ at consistent PN order and performing the  integration  in  \eqref{psiSPA}. Doing so we arrive at 
\begin{align}
	\begin{split}
	\label{ourpsispa}
	\psi_{\rm SPA}(t_f) &  =  \psi_{\rm SPA}^{\rm EH}(t_f)+\psi_{\rm SPA}^{I_1+I_2}(t_f) +\psi_{\rm SPA}^{\rm tidal}(t_f) 
	\  .
	\end{split}
	\end{align}
Here 
\begin{equation}\label{eq:psispaEH}
	\begin{split}
		\psi_{\rm SPA}^{\rm EH}(t_f) &= \ 2 \pi f t_{\rm ref}' - 2 \phi_{\rm ref}' +  \frac{3}{128 \, \nu \,  v_f^5}\, 
	   \end{split}
\end{equation}
is  the EH contribution, where we have also included the reference time and phase $t'_{\rm ref}$ and $\phi'_{\rm ref}$, which have been redefined in order to absorb terms
that depend on $v_{\text{ref}}$; and 
\begin{equation}\label{eq:psispa}
	\begin{split}
		\psi_{\rm SPA}^{I_1+I_2}(t_f) &=	- \frac{3 }{128 \, \nu \,  v_f^5} \left[    156 \frac{\alpha_2 }{(G
	   M)^4}  v_f^{10} \, -\, \frac{545\, \alpha_1+(665-850\, \nu) \ \alpha_2}{14(GM)^4}v_f^{12}\right] \ , 
	   \\
	   \psi_{\rm SPA}^{\rm tidal}(t_f) &=-\frac{3}{128 \, \nu \,  v_f^5} \left\{24\frac{v_f^{10}}{(GM)^4}\left(8(12\lambda_{12}+\lambda_{12}')+12\eta_{12}+\eta_{12}'\right) \right. \\
	   &\hspace{0.5cm} \left.-\frac{10}{7}\frac{v_f^{12}}{(GM)^4}\Big[ 4((91-170\nu)\lambda_{12}-6\lambda_{12}')-5(17\nu + 37)\eta_{12}-9\eta_{12}'\Big]\right\} \, ,
	\end{split} 
\end{equation}
are the new contributions due to cubic and tidal perturbations. 
Similarly to our comment after \eqref{bindingE}, we note that all the terms at leading order in velocity in \eqref{eq:psispa} 
are  complete, while the remaining ones would also receive further modifications from a 3PM computation of the potential and a 2PM computation of the quadrupole. 

Finally, it is interesting to compare our results with those of \cite{Sennett:2019bpc}. The perturbations considered in that paper have the form 
\begin{equation}
\label{L8}
\cL_{8} \,=\, \beta_{1}\, \mathcal{C}^{\, 2}\, +\, \beta_{2}\, \mathcal{C}\, \widetilde{\mathcal{C}}\, +\, \beta_{3}\, \widetilde{\mathcal{C}}^{\, 2}\ ,
\end{equation}  
where 
\begin{equation}
\label{eq:Cs}   
\mathcal{C}  \,:=\, R_{\mu \nu \rho \sigma}\, R^{\mu \nu \rho \sigma}\ , \hspace{2cm} \widetilde{\mathcal{C}} \,:=\dfrac{1}{2}\, R_{\mu \nu \alpha \beta}\, \epsilon^{\alpha \beta}\,_{\gamma \delta}\, R^{\gamma \delta \mu \nu}\ . 
\end{equation}
The  modifications to   $\psi_{\rm SPA}(t_f)$ due to   quartic interactions as found in 
\cite{Sennett:2019bpc}  are  (reinstating powers of $G$ in the result of  that paper, and converting their $d_\Lambda$ into our $\beta_1$ as defined in \eqref{L8}),
\begin{align}
	\begin{split}
	\label{theirpsispa}
	\psi_{\rm SPA}^{\text{quartic}}(t_f)
	   &= \psi_{\rm SPA}^{\rm EH}(t_f)+\frac{3}{128 \, \nu \,  v_f^5} \left[ \left(\frac{234240}{11} - \frac{522240}{11} \nu\right)  \frac{\beta_1}{(GM)^6} v_f^{16} \right]
	   \ . 
	\end{split}
	\end{align} 
Note the different dependence on $v_f$ in the correction terms in \eqref{eq:psispa} and \eqref{theirpsispa}, which are of $\cO (v_f^{10})$ and $\cO (v_f^{16})$ in the leading cubic and  tidal,  and quartic cases, respectively.  
Finally, it will   be interesting to perform a comparison of our  result  in \eqref{ourpsispa} to experimental data, as performed in \cite{Sennett:2019bpc} for the case of quartic perturbations in the Riemann tensor.

\section*{Acknowledgements}

We would like to thank Alessandra Buonanno and  Jung-Wook Kim for very useful  discussions. This work  was supported by the Science and Technology Facilities Council (STFC) Consolidated Grants ST/P000754/1 \textit{``String Theory, Gauge Theory and Duality''} and
ST/T000686/1 \textit{``Amplitudes, Strings and Duality''}, and by the European Union's Horizon 2020 research and innovation programme under the Marie Sk\l{}odowska-Curie grant agreement No.~764850 {\it ``\href{https://sagex.org}{SAGEX}''}.


\appendix

\section{Hamiltonians with momentum-dependent potentials}
\label{app:A}

Consider a momentum-dependent Hamiltonian of the form 
\begin{align}
H = \frac{\vec{p}^{\, 2}}{2\mu} \Big[ 1 + 2 \mu \, U(r)\Big] + V(r)
\ , 
\end{align}
where $\vec{p}^{\, 2} = p_r^2 + \frac{p_{\phi}^2}{r^2}$. From Hamilton's equations we learn that $p_\phi:= l$ is constant, as well as 
$\dot{\phi} = \frac{l}{\mu r^2} \Big[ 1 + 2 \mu U(r)\Big]$.
 The latter equation can be used to re-express $l$ as a function of $\Omega$. 
We also have 
\begin{align}
\dot{r} = \frac{p_r}{\mu } \Big[ 1 + 2 \mu\,  U(r)\Big]\ , 
\end{align}
and, for circular orbits, we see that $p_r=0$ and hence $\dot{p}_r=0$. In this case,   the Hamilton equation $\dot{p}_r = - \frac{\partial H}{\partial r}$ simplifies to
\begin{align}
V^\prime (\ro) - \frac{l^2}{\mu \ro^3} \big[ 1 + 2 \mu U(\ro)\big] + \frac{l^2}{\ro^2} U^\prime (\ro) =0
\ , 
\end{align}
where $\ro$ is the radius of the circular orbit. We will also set $\Omega := \dot{\phi}(r=\ro)$, or 
\begin{align}
\label{Omegatot}
\Omega := \frac{l}{\mu \ro^2} \big[ 1 + 2 \mu U(\ro)\big]\ . 
\end{align}
 Using 
this to eliminate $l$ in favour of $\Omega$, 
we finally get 
\begin{align}\label{eq:vprime}
V^\prime (\ro) - \frac{\mu \ro \Omega^2}{1 + 2 \mu U(\ro)}\Big[ 
 1-  \frac{\mu \ro U^\prime (\ro) }{ 1 + 2 \mu U(\ro)}\Big]  =0
\ . 
\end{align}
This equation determines $\ro$ as a function  of $\Omega$.
In the absence of a perturbation, we have 
\begin{align}
\Omega_{N} = \frac{l}{\mu \rn^2} \ , 
\end{align}
where $r_N$ is the radius of the circular orbit in the EH theory, given in \eqref{erreenne}.

\bibliographystyle{utphys}
\bibliography{remainder}

\providecommand{\href}[2]{#2}\begingroup\raggedright\begin{thebibliography}{10}

\bibitem{Abbott:2016blz}
{\bfseries LIGO Scientific, Virgo} Collaboration, B.~Abbott {\em et~al.},
  ``{Observation of Gravitational Waves from a Binary Black Hole Merger},''
  \href{http://dx.doi.org/10.1103/PhysRevLett.116.061102}{{\em Phys. Rev.
  Lett.} {\bfseries 116} no.~6, (2016) 061102},
  \href{http://arxiv.org/abs/1602.03837}{{\ttfamily arXiv:1602.03837 [gr-qc]}}.

\bibitem{Damour:1985mt}
T.~Damour and G.~Sch{\"a}fer, ``{Lagrangians for point masses at the second
  post-Newtonian approximation of general relativity},''
\href{http://dx.doi.org/10.1007/BF00773685}{{\em Gen. Rel. Grav.} {\bfseries
  17} (1985) 879--905}.

\bibitem{Gilmore:2008gq}
J.~B. Gilmore and A.~Ross, ``{Effective field theory calculation of second
  post-Newtonian binary dynamics},''
  \href{http://dx.doi.org/10.1103/PhysRevD.78.124021}{{\em Phys. Rev.}
  {\bfseries D78} (2008) 124021},
\href{http://arxiv.org/abs/0810.1328}{{\ttfamily arXiv:0810.1328 [gr-qc]}}.

\bibitem{Damour:2001bu}
T.~Damour, P.~Jaranowski, and G.~Sch{\"a}fer, ``{Dimensional regularization of
  the gravitational interaction of point masses},''
  \href{http://dx.doi.org/10.1016/S0370-2693(01)00642-6}{{\em Phys. Lett.}
  {\bfseries B513} (2001) 147--155},
\href{http://arxiv.org/abs/gr-qc/0105038}{{\ttfamily arXiv:gr-qc/0105038
  [gr-qc]}}.

\bibitem{Blanchet:2003gy}
L.~Blanchet, T.~Damour, and G.~Esposito-Farese, ``{Dimensional regularization
  of the third post-Newtonian dynamics of point particles in harmonic
  coordinates},'' \href{http://dx.doi.org/10.1103/PhysRevD.69.124007}{{\em
  Phys. Rev.} {\bfseries D69} (2004) 124007},
\href{http://arxiv.org/abs/gr-qc/0311052}{{\ttfamily arXiv:gr-qc/0311052
  [gr-qc]}}.

\bibitem{Itoh:2003fy}
Y.~Itoh and T.~Futamase, ``{New derivation of a third post-Newtonian equation
  of motion for relativistic compact binaries without ambiguity},''
  \href{http://dx.doi.org/10.1103/PhysRevD.68.121501}{{\em Phys. Rev.}
  {\bfseries D68} (2003) 121501},
\href{http://arxiv.org/abs/gr-qc/0310028}{{\ttfamily arXiv:gr-qc/0310028
  [gr-qc]}}.

\bibitem{Foffa:2011ub}
S.~Foffa and R.~Sturani, ``{Effective field theory calculation of conservative
  binary dynamics at third post-Newtonian order},''
  \href{http://dx.doi.org/10.1103/PhysRevD.84.044031}{{\em Phys. Rev.}
  {\bfseries D84} (2011) 044031},
\href{http://arxiv.org/abs/1104.1122}{{\ttfamily arXiv:1104.1122 [gr-qc]}}.

\bibitem{Jaranowski:2012eb}
P.~Jaranowski and G.~Sch{\"a}fer, ``{Towards the 4th post-Newtonian Hamiltonian
  for two-point-mass systems},''
  \href{http://dx.doi.org/10.1103/PhysRevD.86.061503}{{\em Phys. Rev.}
  {\bfseries D86} (2012) 061503},
\href{http://arxiv.org/abs/1207.5448}{{\ttfamily arXiv:1207.5448 [gr-qc]}}.

\bibitem{Damour:2014jta}
T.~Damour, P.~Jaranowski, and G.~Sch{\"a}fer, ``{Nonlocal-in-time action for
  the fourth post-Newtonian conservative dynamics of two-body systems},''
  \href{http://dx.doi.org/10.1103/PhysRevD.89.064058}{{\em Phys. Rev.}
  {\bfseries D89} no.~6, (2014) 064058},
\href{http://arxiv.org/abs/1401.4548}{{\ttfamily arXiv:1401.4548 [gr-qc]}}.

\bibitem{Galley:2015kus}
C.~R. Galley, A.~K. Leibovich, R.~A. Porto, and A.~Ross, ``{Tail effect in
  gravitational radiation reaction: Time nonlocality and renormalization group
  evolution},'' \href{http://dx.doi.org/10.1103/PhysRevD.93.124010}{{\em Phys.
  Rev. D} {\bfseries 93} (2016) 124010},
  \href{http://arxiv.org/abs/1511.07379}{{\ttfamily arXiv:1511.07379 [gr-qc]}}.

\bibitem{Damour:2015isa}
T.~Damour, P.~Jaranowski, and G.~Sch{\"a}fer, ``{Fourth post-Newtonian
  effective one-body dynamics},''
  \href{http://dx.doi.org/10.1103/PhysRevD.91.084024}{{\em Phys. Rev.}
  {\bfseries D91} no.~8, (2015) 084024},
\href{http://arxiv.org/abs/1502.07245}{{\ttfamily arXiv:1502.07245 [gr-qc]}}.

\bibitem{Damour:2016abl}
T.~Damour, P.~Jaranowski, and G.~Sch{\"a}fer, ``{Conservative dynamics of
  two-body systems at the fourth post-Newtonian approximation of general
  relativity},'' \href{http://dx.doi.org/10.1103/PhysRevD.93.084014}{{\em Phys.
  Rev.} {\bfseries D93} no.~8, (2016) 084014},
\href{http://arxiv.org/abs/1601.01283}{{\ttfamily arXiv:1601.01283 [gr-qc]}}.

\bibitem{Bernard:2015njp}
L.~Bernard, L.~Blanchet, A.~Bohé, G.~Faye, and S.~Marsat, ``{Fokker action of
  nonspinning compact binaries at the fourth post-Newtonian approximation},''
  \href{http://dx.doi.org/10.1103/PhysRevD.93.084037}{{\em Phys. Rev.}
  {\bfseries D93} no.~8, (2016) 084037},
\href{http://arxiv.org/abs/1512.02876}{{\ttfamily arXiv:1512.02876 [gr-qc]}}.

\bibitem{Bernard:2016wrg}
L.~Bernard, L.~Blanchet, A.~Bohé, G.~Faye, and S.~Marsat, ``{Energy and
  periastron advance of compact binaries on circular orbits at the fourth
  post-Newtonian order},''
  \href{http://dx.doi.org/10.1103/PhysRevD.95.044026}{{\em Phys. Rev.}
  {\bfseries D95} no.~4, (2017) 044026},
\href{http://arxiv.org/abs/1610.07934}{{\ttfamily arXiv:1610.07934 [gr-qc]}}.

\bibitem{Foffa:2012rn}
S.~Foffa and R.~Sturani, ``{Dynamics of the gravitational two-body problem at
  fourth post-Newtonian order and at quadratic order in the Newton constant},''
  \href{http://dx.doi.org/10.1103/PhysRevD.87.064011}{{\em Phys. Rev.}
  {\bfseries D87} no.~6, (2013) 064011},
\href{http://arxiv.org/abs/1206.7087}{{\ttfamily arXiv:1206.7087 [gr-qc]}}.

\bibitem{Foffa:2016rgu}
S.~Foffa, P.~Mastrolia, R.~Sturani, and C.~Sturm, ``{Effective field theory
  approach to the gravitational two-body dynamics, at fourth post-Newtonian
  order and quintic in the Newton constant},''
  \href{http://dx.doi.org/10.1103/PhysRevD.95.104009}{{\em Phys. Rev.}
  {\bfseries D95} no.~10, (2017) 104009},
\href{http://arxiv.org/abs/1612.00482}{{\ttfamily arXiv:1612.00482 [gr-qc]}}.

\bibitem{Porto:2017dgs}
R.~A. Porto and I.~Z. Rothstein, ``{Apparent ambiguities in the post-Newtonian
  expansion for binary systems},''
  \href{http://dx.doi.org/10.1103/PhysRevD.96.024062}{{\em Phys. Rev. D}
  {\bfseries 96} no.~2, (2017) 024062},
  \href{http://arxiv.org/abs/1703.06433}{{\ttfamily arXiv:1703.06433 [gr-qc]}}.

\bibitem{Porto:2017shd}
R.~A. Porto, ``{Lamb shift and the gravitational binding energy for binary
  black holes},'' \href{http://dx.doi.org/10.1103/PhysRevD.96.024063}{{\em
  Phys. Rev. D} {\bfseries 96} no.~2, (2017) 024063},
  \href{http://arxiv.org/abs/1703.06434}{{\ttfamily arXiv:1703.06434 [gr-qc]}}.

\bibitem{Foffa:2019yfl}
S.~Foffa, R.~A. Porto, I.~Rothstein, and R.~Sturani, ``{Conservative dynamics
  of binary systems to fourth Post-Newtonian order in the EFT approach II:
  Renormalized Lagrangian},''
  \href{http://dx.doi.org/10.1103/PhysRevD.100.024048}{{\em Phys. Rev. D}
  {\bfseries 100} no.~2, (2019) 024048},
  \href{http://arxiv.org/abs/1903.05118}{{\ttfamily arXiv:1903.05118 [gr-qc]}}.

\bibitem{Blumlein:2020pog}
J.~Bluemlein, A.~Maier, P.~Marquard, and G.~Schaefer, ``{Fourth post-Newtonian
  Hamiltonian dynamics of two-body systems from an effective field theory
  approach},'' \href{http://dx.doi.org/10.1016/j.nuclphysb.2020.115041}{{\em
  Nucl. Phys. B} {\bfseries 955} (2020) 115041},
  \href{http://arxiv.org/abs/2003.01692}{{\ttfamily arXiv:2003.01692 [gr-qc]}}.

\bibitem{Foffa:2019hrb}
S.~Foffa, P.~Mastrolia, R.~Sturani, C.~Sturm, and W.~J. Torres~Bobadilla,
  ``{Static two-body potential at fifth post-Newtonian order},''
  \href{http://dx.doi.org/10.1103/PhysRevLett.122.241605}{{\em Phys. Rev.
  Lett.} {\bfseries 122} no.~24, (2019) 241605},
  \href{http://arxiv.org/abs/1902.10571}{{\ttfamily arXiv:1902.10571 [gr-qc]}}.

\bibitem{Blumlein:2019zku}
J.~Bluemlein, A.~Maier, and P.~Marquard, ``{Five-Loop Static Contribution to
  the Gravitational Interaction Potential of Two Point Masses},''
  \href{http://dx.doi.org/10.1016/j.physletb.2019.135100}{{\em Phys. Lett. B}
  {\bfseries 800} (2020) 135100},
  \href{http://arxiv.org/abs/1902.11180}{{\ttfamily arXiv:1902.11180 [gr-qc]}}.

\bibitem{Blumlein:2020pyo}
J.~Bl\"umlein, A.~Maier, P.~Marquard, and G.~Sch\"afer, ``{The fifth-order
  post-Newtonian Hamiltonian dynamics of two-body systems from an effective
  field theory approach: potential contributions},''
  \href{http://arxiv.org/abs/2010.13672}{{\ttfamily arXiv:2010.13672 [gr-qc]}}.

\bibitem{Blumlein:2020znm}
J.~Bluemlein, A.~Maier, P.~Marquard, and G.~Schaefer, ``{Testing binary
  dynamics in gravity at the sixth post-Newtonian level},''
  \href{http://arxiv.org/abs/2003.07145}{{\ttfamily arXiv:2003.07145 [gr-qc]}}.

\bibitem{Bini:2020uiq}
D.~Bini, T.~Damour, A.~Geralico, S.~Laporta, and P.~Mastrolia, ``{Gravitational
  dynamics at $O(G^6)$: perturbative gravitational scattering meets
  experimental mathematics},''
  \href{http://arxiv.org/abs/2008.09389}{{\ttfamily arXiv:2008.09389 [gr-qc]}}.

\bibitem{Goldberger:2004jt}
W.~D. Goldberger and I.~Z. Rothstein, ``{An Effective field theory of gravity
  for extended objects},''
  \href{http://dx.doi.org/10.1103/PhysRevD.73.104029}{{\em Phys. Rev.}
  {\bfseries D73} (2006) 104029},
\href{http://arxiv.org/abs/hep-th/0409156}{{\ttfamily arXiv:hep-th/0409156
  [hep-th]}}.

\bibitem{Porto:2016pyg}
R.~A. Porto, ``{The effective field theorist's approach to gravitational
  dynamics},'' \href{http://dx.doi.org/10.1016/j.physrep.2016.04.003}{{\em
  Phys. Rept.} {\bfseries 633} (2016) 1--104},
  \href{http://arxiv.org/abs/1601.04914}{{\ttfamily arXiv:1601.04914
  [hep-th]}}.

\bibitem{Bern:2019nnu}
Z.~Bern, C.~Cheung, R.~Roiban, C.-H. Shen, M.~P. Solon, and M.~Zeng,
  ``{Scattering Amplitudes and the Conservative Hamiltonian for Binary Systems
  at Third Post-Minkowskian Order},''
  \href{http://dx.doi.org/10.1103/PhysRevLett.122.201603}{{\em Phys. Rev.
  Lett.} {\bfseries 122} no.~20, (2019) 201603},
  \href{http://arxiv.org/abs/1901.04424}{{\ttfamily arXiv:1901.04424
  [hep-th]}}.

\bibitem{Bern:2019crd}
Z.~Bern, C.~Cheung, R.~Roiban, C.-H. Shen, M.~P. Solon, and M.~Zeng, ``{Black
  Hole Binary Dynamics from the Double Copy and Effective Theory},''
  \href{http://dx.doi.org/10.1007/JHEP10(2019)206}{{\em JHEP} {\bfseries 10}
  (2019) 206}, \href{http://arxiv.org/abs/1908.01493}{{\ttfamily
  arXiv:1908.01493 [hep-th]}}.

\bibitem{Cheung:2020gyp}
C.~Cheung and M.~P. Solon, ``{Classical Gravitational Scattering at ${\cal
  O}(G^3)$ from Feynman Diagrams},''
  \href{http://arxiv.org/abs/2003.08351}{{\ttfamily arXiv:2003.08351
  [hep-th]}}.

\bibitem{Bjerrum-Bohr:2013bxa}
N.~E.~J. Bjerrum-Bohr, J.~F. Donoghue, and P.~Vanhove, ``{On-shell Techniques
  and Universal Results in Quantum Gravity},''
  \href{http://dx.doi.org/10.1007/JHEP02(2014)111}{{\em JHEP} {\bfseries 02}
  (2014) 111},
\href{http://arxiv.org/abs/1309.0804}{{\ttfamily arXiv:1309.0804 [hep-th]}}.

\bibitem{Bjerrum-Bohr:2017dxw}
N.~E.~J. Bjerrum-Bohr, B.~R. Holstein, J.~F. Donoghue, L.~Plante, and
  P.~Vanhove, ``{Illuminating Light Bending},''
  \href{http://dx.doi.org/10.22323/1.292.0077}{{\em PoS} {\bfseries CORFU2016}
  (2017) 077},
\href{http://arxiv.org/abs/1704.01624}{{\ttfamily arXiv:1704.01624 [gr-qc]}}.

\bibitem{Luna:2017dtq}
A.~Luna, I.~Nicholson, D.~O'Connell, and C.~D. White, ``{Inelastic Black Hole
  Scattering from Charged Scalar Amplitudes},''
  \href{http://dx.doi.org/10.1007/JHEP03(2018)044}{{\em JHEP} {\bfseries 03}
  (2018) 044}, \href{http://arxiv.org/abs/1711.03901}{{\ttfamily
  arXiv:1711.03901 [hep-th]}}.

\bibitem{Kosower:2018adc}
D.~A. Kosower, B.~Maybee, and D.~O'Connell, ``{Amplitudes, Observables, and
  Classical Scattering},''
  \href{http://dx.doi.org/10.1007/JHEP02(2019)137}{{\em JHEP} {\bfseries 02}
  (2019) 137},
\href{http://arxiv.org/abs/1811.10950}{{\ttfamily arXiv:1811.10950 [hep-th]}}.

\bibitem{Guevara:2018wpp}
A.~Guevara, A.~Ochirov, and J.~Vines, ``{Scattering of Spinning Black Holes
  from Exponentiated Soft Factors},''
\href{http://arxiv.org/abs/1812.06895}{{\ttfamily arXiv:1812.06895 [hep-th]}}.

\bibitem{Chung:2018kqs}
M.-Z. Chung, Y.-T. Huang, J.-W. Kim, and S.~Lee, ``{The simplest massive
  S-matrix: from minimal coupling to Black Holes},''
  \href{http://dx.doi.org/10.1007/JHEP04(2019)156}{{\em JHEP} {\bfseries 04}
  (2019) 156}, \href{http://arxiv.org/abs/1812.08752}{{\ttfamily
  arXiv:1812.08752 [hep-th]}}.

\bibitem{KoemansCollado:2019ggb}
A.~Koemans~Collado, P.~Di~Vecchia, and R.~Russo, ``{Revisiting the 2PM eikonal
  and the dynamics of binary black holes},''
\href{http://arxiv.org/abs/1904.02667}{{\ttfamily arXiv:1904.02667 [hep-th]}}.

\bibitem{Maybee:2019jus}
B.~Maybee, D.~O'Connell, and J.~Vines, ``{Observables and amplitudes for
  spinning particles and black holes},''
\href{http://arxiv.org/abs/1906.09260}{{\ttfamily arXiv:1906.09260 [hep-th]}}.

\bibitem{Guevara:2019fsj}
A.~Guevara, A.~Ochirov, and J.~Vines, ``{Black-hole scattering with general
  spin directions from minimal-coupling amplitudes},''
  \href{http://dx.doi.org/10.1103/PhysRevD.100.104024}{{\em Phys. Rev.}
  {\bfseries D100} (2019) 104024},
\href{http://arxiv.org/abs/1906.10071}{{\ttfamily arXiv:1906.10071 [hep-th]}}.

\bibitem{Chung:2019duq}
M.-Z. Chung, Y.-T. Huang, and J.-W. Kim, ``{Classical potential for general
  spinning bodies},'' \href{http://dx.doi.org/10.1007/JHEP09(2020)074}{{\em
  JHEP} {\bfseries 09} (2020) 074},
  \href{http://arxiv.org/abs/1908.08463}{{\ttfamily arXiv:1908.08463
  [hep-th]}}.

\bibitem{Damgaard:2019lfh}
P.~H. Damgaard, K.~Haddad, and A.~Helset, ``{Heavy Black Hole Effective
  Theory},'' \href{http://dx.doi.org/10.1007/JHEP11(2019)070}{{\em JHEP}
  {\bfseries 11} (2019) 070}, \href{http://arxiv.org/abs/1908.10308}{{\ttfamily
  arXiv:1908.10308 [hep-ph]}}.

\bibitem{Kalin:2019rwq}
G.~K\"alin and R.~A. Porto, ``{From Boundary Data to Bound States},''
  \href{http://dx.doi.org/10.1007/JHEP01(2020)072}{{\em JHEP} {\bfseries 01}
  (2020) 072}, \href{http://arxiv.org/abs/1910.03008}{{\ttfamily
  arXiv:1910.03008 [hep-th]}}.

\bibitem{Kalin:2019inp}
G.~K\"alin and R.~A. Porto, ``{From boundary data to bound states. Part II.
  Scattering angle to dynamical invariants (with twist)},''
  \href{http://dx.doi.org/10.1007/JHEP02(2020)120}{{\em JHEP} {\bfseries 02}
  (2020) 120}, \href{http://arxiv.org/abs/1911.09130}{{\ttfamily
  arXiv:1911.09130 [hep-th]}}.

\bibitem{Chung:2020rrz}
M.-Z. Chung, Y.-t. Huang, J.-W. Kim, and S.~Lee, ``{Complete Hamiltonian for
  spinning binary systems at first post-Minkowskian order},''
  \href{http://dx.doi.org/10.1007/JHEP05(2020)105}{{\em JHEP} {\bfseries 05}
  (2020) 105}, \href{http://arxiv.org/abs/2003.06600}{{\ttfamily
  arXiv:2003.06600 [hep-th]}}.

\bibitem{Cristofoli:2020uzm}
A.~Cristofoli, P.~H. Damgaard, P.~Di~Vecchia, and C.~Heissenberg,
  ``{Second-order Post-Minkowskian scattering in arbitrary dimensions},''
  \href{http://dx.doi.org/10.1007/JHEP07(2020)122}{{\em JHEP} {\bfseries 07}
  (2020) 122}, \href{http://arxiv.org/abs/2003.10274}{{\ttfamily
  arXiv:2003.10274 [hep-th]}}.

\bibitem{Bern:2020gjj}
Z.~Bern, H.~Ita, J.~Parra-Martinez, and M.~S. Ruf, ``{Universality in the
  classical limit of massless gravitational scattering},''
  \href{http://arxiv.org/abs/2002.02459}{{\ttfamily arXiv:2002.02459
  [hep-th]}}.

\bibitem{Bern:2020buy}
Z.~Bern, A.~Luna, R.~Roiban, C.-H. Shen, and M.~Zeng, ``{Spinning Black Hole
  Binary Dynamics, Scattering Amplitudes and Effective Field Theory},''
  \href{http://arxiv.org/abs/2005.03071}{{\ttfamily arXiv:2005.03071
  [hep-th]}}.

\bibitem{Parra-Martinez:2020dzs}
J.~Parra-Martinez, M.~S. Ruf, and M.~Zeng, ``{Extremal black hole scattering at
  $O(G^3)$: graviton dominance, eikonal exponentiation, and differential
  equations},'' \href{http://arxiv.org/abs/2005.04236}{{\ttfamily
  arXiv:2005.04236 [hep-th]}}.

\bibitem{delaCruz:2020bbn}
L.~de~la Cruz, B.~Maybee, D.~O'Connell, and A.~Ross, ``{Classical Yang-Mills
  observables from amplitudes},''
  \href{http://arxiv.org/abs/2009.03842}{{\ttfamily arXiv:2009.03842
  [hep-th]}}.

\bibitem{Emond:2020lwi}
W.~T. Emond, Y.-T. Huang, U.~Kol, N.~Moynihan, and D.~O'Connell, ``{Amplitudes
  from Coulomb to Kerr-Taub-NUT},''
  \href{http://arxiv.org/abs/2010.07861}{{\ttfamily arXiv:2010.07861
  [hep-th]}}.

\bibitem{Donoghue:1994dn}
J.~F. Donoghue, ``{General relativity as an effective field theory: The leading
  quantum corrections},''
  \href{http://dx.doi.org/10.1103/PhysRevD.50.3874}{{\em Phys. Rev.} {\bfseries
  D50} (1994) 3874--3888},
\href{http://arxiv.org/abs/gr-qc/9405057}{{\ttfamily arXiv:gr-qc/9405057
  [gr-qc]}}.

\bibitem{Endlich:2017tqa}
S.~Endlich, V.~Gorbenko, J.~Huang, and L.~Senatore, ``{An effective formalism
  for testing extensions to General Relativity with gravitational waves},''
  \href{http://dx.doi.org/10.1007/JHEP09(2017)122}{{\em JHEP} {\bfseries 09}
  (2017) 122}, \href{http://arxiv.org/abs/1704.01590}{{\ttfamily
  arXiv:1704.01590 [gr-qc]}}.

\bibitem{Sennett:2019bpc}
N.~Sennett, R.~Brito, A.~Buonanno, V.~Gorbenko, and L.~Senatore,
  ``{Gravitational-Wave Constraints on an Effective--Field-Theory Extension of
  General Relativity},'' \href{http://arxiv.org/abs/1912.09917}{{\ttfamily
  arXiv:1912.09917 [gr-qc]}}.

\bibitem{Brandhuber:2019qpg}
A.~Brandhuber and G.~Travaglini, ``{On higher-derivative effects on the
  gravitational potential and particle bending},''
  \href{http://dx.doi.org/10.1007/JHEP01(2020)010}{{\em JHEP} {\bfseries 01}
  (2020) 010},
\href{http://arxiv.org/abs/1905.05657}{{\ttfamily arXiv:1905.05657 [hep-th]}}.

\bibitem{Emond:2019crr}
W.~T. Emond and N.~Moynihan, ``{Scattering Amplitudes, Black Holes and Leading
  Singularities in Cubic Theories of Gravity},''
  \href{http://dx.doi.org/10.1007/JHEP12(2019)019}{{\em JHEP} {\bfseries 12}
  (2019) 019}, \href{http://arxiv.org/abs/1905.08213}{{\ttfamily
  arXiv:1905.08213 [hep-th]}}.

\bibitem{AccettulliHuber:2020oou}
M.~Accettulli~Huber, A.~Brandhuber, S.~De~Angelis, and G.~Travaglini,
  ``{Eikonal phase matrix, deflection angle and time delay in effective field
  theories of gravity},''
  \href{http://dx.doi.org/10.1103/PhysRevD.102.046014}{{\em Phys. Rev. D}
  {\bfseries 102} no.~4, (2020) 046014},
  \href{http://arxiv.org/abs/2006.02375}{{\ttfamily arXiv:2006.02375
  [hep-th]}}.

\bibitem{Huber:2019ugz}
M.~Accettulli~Huber, A.~Brandhuber, S.~De~Angelis, and G.~Travaglini, ``{Note
  on the absence of $R^2$ corrections to Newton's potential},''
  \href{http://dx.doi.org/10.1103/PhysRevD.101.046011}{{\em Phys. Rev.}
  {\bfseries D101} no.~4, (2020) 046011},
\href{http://arxiv.org/abs/1911.10108}{{\ttfamily arXiv:1911.10108 [hep-th]}}.

\bibitem{deRham:2019ctd}
C.~de~Rham and A.~J. Tolley, ``{Speed of gravity},''
  \href{http://dx.doi.org/10.1103/PhysRevD.101.063518}{{\em Phys. Rev. D}
  {\bfseries 101} no.~6, (2020) 063518},
  \href{http://arxiv.org/abs/1909.00881}{{\ttfamily arXiv:1909.00881
  [hep-th]}}.

\bibitem{Bern:2020uwk}
Z.~Bern, J.~Parra-Martinez, R.~Roiban, E.~Sawyer, and C.-H. Shen, ``{Leading
  Nonlinear Tidal Effects and Scattering Amplitudes},''
  \href{http://arxiv.org/abs/2010.08559}{{\ttfamily arXiv:2010.08559
  [hep-th]}}.

\bibitem{Avramidi:1986mj}
I.~G. Avramidi, {\em {Covariant methods for the calculation of the effective
  action in quantum field theory and investigation of higher derivative quantum
  gravity}}.
\newblock PhD thesis, Moscow State U., 1986.
\newblock
\href{http://arxiv.org/abs/hep-th/9510140}{{\ttfamily arXiv:hep-th/9510140
  [hep-th]}}.
\newblock

\bibitem{Avramidi:1990je}
I.~G. Avramidi, ``{The Covariant Technique for Calculation of One Loop
  Effective Action},'' \href{http://dx.doi.org/10.1016/0550-3213(91)90492-G,
  10.1016/S0550-3213(97)00717-7}{{\em Nucl. Phys.} {\bfseries B355} (1991)
  712--754}.
[Erratum: Nucl. Phys.B509,557(1998)].

\bibitem{vanNieuwenhuizen:1976vb}
P.~van Nieuwenhuizen and C.~C. Wu, ``{On Integral Relations for Invariants
  Constructed from Three Riemann Tensors and their Applications in Quantum
  Gravity},''
\href{http://dx.doi.org/10.1063/1.523128}{{\em J. Math. Phys.} {\bfseries 18}
  (1977) 182}.

\bibitem{Camanho:2014apa}
X.~O. Camanho, J.~D. Edelstein, J.~Maldacena, and A.~Zhiboedov, ``{Causality
  Constraints on Corrections to the Graviton Three-Point Coupling},''
  \href{http://dx.doi.org/10.1007/JHEP02(2016)020}{{\em JHEP} {\bfseries 02}
  (2016) 020},
\href{http://arxiv.org/abs/1407.5597}{{\ttfamily arXiv:1407.5597 [hep-th]}}.

\bibitem{deRham:2020ejn}
C.~de~Rham, J.~Francfort, and J.~Zhang, ``{Black Hole Gravitational Waves in
  the Effective Field Theory of Gravity},''
  \href{http://dx.doi.org/10.1103/PhysRevD.102.024079}{{\em Phys. Rev. D}
  {\bfseries 102} no.~2, (2020) 024079},
  \href{http://arxiv.org/abs/2005.13923}{{\ttfamily arXiv:2005.13923
  [hep-th]}}.

\bibitem{deRham:2020zyh}
C.~de~Rham and A.~J. Tolley, ``{Causality in curved spacetimes: The speed of
  light and gravity},''
  \href{http://dx.doi.org/10.1103/PhysRevD.102.084048}{{\em Phys. Rev. D}
  {\bfseries 102} no.~8, (2020) 084048},
  \href{http://arxiv.org/abs/2007.01847}{{\ttfamily arXiv:2007.01847
  [hep-th]}}.

\bibitem{Drummond:1979pp}
I.~Drummond and S.~Hathrell, ``{QED Vacuum Polarization in a Background
  Gravitational Field and Its Effect on the Velocity of Photons},''
  \href{http://dx.doi.org/10.1103/PhysRevD.22.343}{{\em Phys. Rev. D}
  {\bfseries 22} (1980) 343}.

\bibitem{Hollowood:2007ku}
T.~J. Hollowood and G.~M. Shore, ``{The Refractive index of curved spacetime:
  The Fate of causality in QED},''
  \href{http://dx.doi.org/10.1016/j.nuclphysb.2007.11.034}{{\em Nucl. Phys. B}
  {\bfseries 795} (2008) 138--171},
  \href{http://arxiv.org/abs/0707.2303}{{\ttfamily arXiv:0707.2303 [hep-th]}}.

\bibitem{Goon:2016une}
G.~Goon and K.~Hinterbichler, ``{Superluminality, black holes and EFT},''
  \href{http://dx.doi.org/10.1007/JHEP02(2017)134}{{\em JHEP} {\bfseries 02}
  (2017) 134}, \href{http://arxiv.org/abs/1609.00723}{{\ttfamily
  arXiv:1609.00723 [hep-th]}}.

\bibitem{Henry:2019xhg}
Q.~Henry, G.~Faye, and L.~Blanchet, ``{Tidal effects in the equations of motion
  of compact binary systems to next-to-next-to-leading post-Newtonian order},''
  \href{http://dx.doi.org/10.1103/PhysRevD.101.064047}{{\em Phys. Rev. D}
  {\bfseries 101} no.~6, (2020) 064047},
  \href{http://arxiv.org/abs/1912.01920}{{\ttfamily arXiv:1912.01920 [gr-qc]}}.

\bibitem{Marchand:2020fpt}
T.~Marchand, Q.~Henry, F.~Larrouturou, S.~Marsat, G.~Faye, and L.~Blanchet,
  ``{The mass quadrupole moment of compact binary systems at the fourth
  post-Newtonian order},''
  \href{http://dx.doi.org/10.1088/1361-6382/ab9ce1}{{\em Class. Quant. Grav.}
  {\bfseries 37} no.~21, (2020) 215006},
  \href{http://arxiv.org/abs/2003.13672}{{\ttfamily arXiv:2003.13672 [gr-qc]}}.

\bibitem{Henry:2020ski}
Q.~Henry, G.~Faye, and L.~Blanchet, ``{Tidal effects in the gravitational-wave
  phase evolution of compact binary systems to next-to-next-to-leading
  post-Newtonian order},''
  \href{http://dx.doi.org/10.1103/PhysRevD.102.044033}{{\em Phys. Rev. D}
  {\bfseries 102} no.~4, (2020) 044033},
  \href{http://arxiv.org/abs/2005.13367}{{\ttfamily arXiv:2005.13367 [gr-qc]}}.

\bibitem{Damour:1997ub}
T.~Damour, B.~R. Iyer, and B.~Sathyaprakash, ``{Improved filters for
  gravitational waves from inspiralling compact binaries},''
  \href{http://dx.doi.org/10.1103/PhysRevD.57.885}{{\em Phys. Rev. D}
  {\bfseries 57} (1998) 885--907},
  \href{http://arxiv.org/abs/gr-qc/9708034}{{\ttfamily arXiv:gr-qc/9708034}}.

\bibitem{Buonanno:2009zt}
A.~Buonanno, B.~Iyer, E.~Ochsner, Y.~Pan, and B.~Sathyaprakash, ``{Comparison
  of post-Newtonian templates for compact binary inspiral signals in
  gravitational-wave detectors},''
  \href{http://dx.doi.org/10.1103/PhysRevD.80.084043}{{\em Phys. Rev. D}
  {\bfseries 80} (2009) 084043},
  \href{http://arxiv.org/abs/0907.0700}{{\ttfamily arXiv:0907.0700 [gr-qc]}}.

\bibitem{Bini:2020flp}
D.~Bini, T.~Damour, and A.~Geralico, ``{Scattering of tidally interacting
  bodies in post-Minkowskian gravity},''
  \href{http://dx.doi.org/10.1103/PhysRevD.101.044039}{{\em Phys. Rev. D}
  {\bfseries 101} no.~4, (2020) 044039},
  \href{http://arxiv.org/abs/2001.00352}{{\ttfamily arXiv:2001.00352 [gr-qc]}}.

\bibitem{Kalin:2020mvi}
G.~Kälin and R.~A. Porto, ``{Post-Minkowskian Effective Field Theory for
  Conservative Binary Dynamics},''
  \href{http://arxiv.org/abs/2006.01184}{{\ttfamily arXiv:2006.01184
  [hep-th]}}.

\bibitem{Cheung:2020sdj}
C.~Cheung and M.~P. Solon, ``{Tidal Effects in the Post-Minkowskian
  Expansion},'' \href{http://arxiv.org/abs/2006.06665}{{\ttfamily
  arXiv:2006.06665 [hep-th]}}.

\bibitem{Haddad:2020que}
K.~Haddad and A.~Helset, ``{Gravitational tidal effects in quantum field
  theory},'' \href{http://arxiv.org/abs/2008.04920}{{\ttfamily arXiv:2008.04920
  [hep-th]}}.

\bibitem{lovelock_1970}
D.~Lovelock, ``Dimensionally dependent identities,''
  \href{http://dx.doi.org/10.1017/S0305004100046144}{{\em Mathematical
  Proceedings of the Cambridge Philosophical Society} {\bfseries 68} no.~2,
  (1970) 345?350}.

\bibitem{Edgar:2001vv}
S.~Edgar and A.~Hoglund, ``{Dimensionally dependent tensor identities by double
  antisymmetrization},'' \href{http://dx.doi.org/10.1063/1.1425428}{{\em J.
  Math. Phys.} {\bfseries 43} (2002) 659--677},
  \href{http://arxiv.org/abs/gr-qc/0105066}{{\ttfamily arXiv:gr-qc/0105066}}.

\bibitem{Broedel:2012rc}
J.~Broedel and L.~J. Dixon, ``{Color-kinematics duality and double-copy
  construction for amplitudes from higher-dimension operators},''
  \href{http://dx.doi.org/10.1007/JHEP10(2012)091}{{\em JHEP} {\bfseries 10}
  (2012) 091},
\href{http://arxiv.org/abs/1208.0876}{{\ttfamily arXiv:1208.0876 [hep-th]}}.

\bibitem{Cardoso:2018ptl}
V.~Cardoso, M.~Kimura, A.~Maselli, and L.~Senatore, ``{Black Holes in an
  Effective Field Theory Extension of General Relativity},''
  \href{http://dx.doi.org/10.1103/PhysRevLett.121.251105}{{\em Phys. Rev.
  Lett.} {\bfseries 121} no.~25, (2018) 251105},
  \href{http://arxiv.org/abs/1808.08962}{{\ttfamily arXiv:1808.08962 [gr-qc]}}.

\bibitem{Cai:2019npx}
S.~Cai and K.-D. Wang, ``{Non-vanishing of tidal Love numbers},''
  \href{http://arxiv.org/abs/1906.06850}{{\ttfamily arXiv:1906.06850
  [hep-th]}}.

\bibitem{Xu:1987}
X.~Dianyan, ``Two important invariant identities,''
  \href{http://dx.doi.org/10.1103/PhysRevD.35.769}{{\em Phys. Rev. D}
  {\bfseries 35} (Jan, 1987) 769--770}.
  \url{https://link.aps.org/doi/10.1103/PhysRevD.35.769}.

\bibitem{Kim:2020dif}
J.-W. Kim and M.~Shim, ``{Sum rule for Love},''
  \href{http://arxiv.org/abs/2011.03337}{{\ttfamily arXiv:2011.03337
  [hep-th]}}.

\end{thebibliography}\endgroup

\end{document}